\documentclass[global,twocolumn]{svjour}
\usepackage{graphics}
\usepackage{epstopdf}
\usepackage{subfigure}
\usepackage{amsmath}
\usepackage{dsfont}
\journalname{myjournal}
\begin{document}
\title{$T^3$-interferometer for atoms\thanks{Our proposal of an atom interferometer with enhanced phase sensitivity
was inspired by the seminal experiment of H\"ansch and collaborators \cite{Haensch Weitz}
to test the equivalence principle of general relativity based on a matter-wave interferometer for two different isotopes of rubidium.
Whereas in Ref. \cite{Haensch Weitz} the increase in phase sensitivity
originates from quadratic phases reminiscent of the Talbot effect we employ the cubic phase of the quantum mechanical propagator
for a particle moving in a linear potential combined with a Ramsey interferometer.
It is with great pleasure that we dedicate this article to Theodor W. H\"ansch on the occasion of his 75$^{\rm th}$ birthday.}
}
\author{M. Zimmermann\inst{1}, M.A. Efremov\inst{1}, A. Roura\inst{1}, W.P. Schleich\inst{1,2}, S.A. DeSavage\inst{3}, J.P. Davis\inst{4},\\
A. Srinivasan\inst{5}, F.A. Narducci\inst{3}, S.A. Werner\inst{6}, and E.M. Rasel\inst{7}
}                     
%
%
\institute{Institut f\"ur Quantenphysik and Center for Integrated Quantum Science and Technology (IQ$^{ST}$),
Universit\"at Ulm, D-89081 Ulm, Germany\and
Texas A$\&$M University Institute for Advanced Study (TIAS), Institute for Quantum Science and Engineering
(IQSE) and Department of Physics and Astronomy, Texas A$\&$M University, College
Station, Texas 77843-4242, USA\and
Naval Air Systems Command, EO Sensors Division, Patuxent River, Maryland 20670, USA\and
AMPAC, North Wales, Pennsylvania 19154, USA\and
St. Mary's College of Maryland, St. Mary's City, Maryland 20686-3001, USA\and
Physics Laboratory, NIST, Gaithersburg, Maryland 20899, USA\and
Institut f\"ur Quantenoptik, Leibniz Universit\"at Hannover, D-30167 Hannover, Germany}
\date{Received: \today / Revised version: date}
%
\authorrunning{M. Zimmermann et al.}
\maketitle

\begin{abstract}

The quantum mechanical propagator of a massive particle in a linear gravitational potential
derived already in 1927 by Earle H. Kennard \cite{Kennard,Kennard2} contains a phase that scales with the third power of the time $T$
during which the particle experiences the corresponding force.
Since in conventional atom interferometers the internal atomic states are all exposed to the same acceleration $a$,
this $T^3$-phase cancels out and the interferometer phase scales as $T^2$.
In contrast, by applying an external magnetic field we prepare two different accelerations $a_1$ and $a_2$
for two internal states of the atom, which translate themselves into two different cubic phases and
the resulting interferometer phase scales as $T^3$. We present the theoretical background for, and summarize our progress
towards experimentally realizing such a novel atom interferometer.
\end{abstract}

\section{Introduction}
\label{intro}

Phases play an extraordinary role in quantum theory. On one hand, they represent the central
ingredient of wave mechanics $\rm \grave{a}$ la Schr\"odinger,
and on the other, they build a bridge to classical mechanics $\rm \grave{a}$ la Hamilton-Jacobi \cite{Scully}.
For these reasons they constitute a crucial ingredient of matter wave interferometers \cite{Berman,RMP,Fermi_school}
which nowadays represent standard tools for precision measurements.
In this article we consider a new type of atom interferometer whose phase scales with the cube of the time $T$ 
the atom spends in the interferometer.

\subsection{At the interface of quantum and gravity}

During an impromptu seminar at the NATO Advanced Summer Institute at Bad Windsheim in 1981 Eugene Paul Wigner expressed his discomfort
with general relativity \cite{Misner} in view of quantum mechanics \cite{Bohm-QM}. He emphasized that the notion of a space-time trajectory which is a crucial element
of gravity is incompatible with the uncertainty relation of quantum mechanics. Guided by the work of Niels Bohr and Leon Rosenfeld
on the limitations of the electromagnetic field \cite{Bohr-Rosenfeld} due to quantum fluctuations he argued that quantum theory puts severe restrictions
on the measurement of the metric tensor representing the gravitational field.

Wigner's thoughts expressed in this seminar were a consequence of his work several decades earlier. Indeed, already in 1958 together with
Helmut Salecker he had constructed a clock \cite{Salecker-Wigner} based on the reflection of light signals from two mirrors and had analyzed the restrictions
of the uncertainty relation on the weight of the mirrors.\footnote[1]{John Archibald Wheeler frequently emphasized in conversations
about this topic and in print \cite{Wheeler} that these estimates were too conservative. However, to the best of our knowledge they have never been improved.}

In this context it is also worth mentioning that H. Salecker at the {\it Conference on the Role of Gravitation in Physics}
in Chapel Hill in 1957 triggered \cite{DeWitt} a discussion on the equivalence principle by a gedanken experiment involving
a stream of particles being scattered off a diffraction grating.
Daniel M. Greenberger \cite{Greenberger-JMathPhys} a decade later considered a similar arrangement and 
even argued that mass in quantum mechanics should be an operator.

The celebrated Colella-Overhauser-Werner (COW) experiment \cite{COW-experiment} performed in 1975 propelled these and many other
thoughts about the interference of quantum and gravity to the real world. Indeed, based on the de Broglie wave nature of neutrons \cite{book_Werner}
the COW experiment could measure for the first time the phase shift between two arms of a neutron interferometer induced by the gravitational
potential of the earth \cite{COW-experiment,Werner_Gravity_inertia}.

The development of new sources of cold atoms \cite{RMP} as well as molecules \cite{Arndt} and, in particular,
the realization of Bose-Einstein condensates \cite{BEC-Nobel Lectures}
has ushered in a new era of experiments at the interface of quantum mechanics and gravity. Now, novel tests of the equivalence principle
based on matter wave interferometry of different isotopes of the same atom \cite{Haensch Weitz} such as $^{85}{\rm Rb}$ and $^{87}{\rm Rb}$
or even different species \cite{Schlippert} such as $^{39}{\rm K}$ and $^{87}{\rm Rb}$ could be performed.
Even the detection of gravitational waves based on atom interferometry is pursued today \cite{Kasevich}.
Recently Ref. \cite{Pikovski} suggested that gravitational decoherence gives rise to a universal decoherence.
Moreover, the atom laser \cite{Koehl-Haenech} utilizes gravity to form an Airy mode.
Furthermore, it is mind-boggling that nowadays measurements of atomic transitions \cite{Wineland,Matveev-Haenech} 
are sensitive to the redshift of the gravitational field.

An extremely interesting development in this realm is again taking place in the field of neutron optics due to the experimental realization
of the quantum bouncer \cite{Abele}. Here neutrons exposed to the gravitational field of the earth are reflected from a surface and oscillate up and down.
In particular, they experience a potential consisting of the linear ramp and an infinitely steep wall.
It is amazing that a measurement of the transitions between the resulting discrete energy levels 
can put upper bounds \cite{Abele-PRL} on dark energy and dark matter scenarios.

\subsection{Drive for enhanced sensitivity}

Hopefully these few examples convey the excitement in this field of quantum optics and gravity.
Notwithstanding the fact that we still do not have a complete understanding of quantum gravity \cite{Kiefer} we have come a long way
since the early Salecker-Wigner discussions but many questions remain. Indeed, goals such as gravitational wave detection or
a compact gravimeter \cite{Atom-chip-gravimeter} based on atom optics drive the strive for higher sensitivity of these devices.

The Kasevich-Chu atom interferometer \cite{Kasevich-Chu-1,Kasevich-Chu-2}
which is the work horse of atom optics is analogous to the neutron interferometer employed by the COW experiment.
The Bragg diffraction of the neutron from three crystal planes of a silicon slab are replaced by Raman diffraction of the atom
from three standing light crystals. As a result, the phase shift introduced by the gravitational potential is {\it quadratic} in the time $2T$
the particle spends in the interferometer.

Needless to say a different scaling of the phase shift, for example with $T^3$, could lead to an enhanced sensitivity of the interferometer.
In the present article we propose such an interferometer and describe our progress towards realizing it.

Our device rests on three principles: (i) We take advantage of the cubic dependence of the phase in the propagator in a linear potential.
(ii) We employ two different internal states of the atom which experience different accelerations.
(iii) We close the interferometer in position {\it and} velocity by a sequence of four laser pulses.

\begin{figure*}
\centering
\subfigure[]{
\resizebox{!}{5.5cm}{\includegraphics{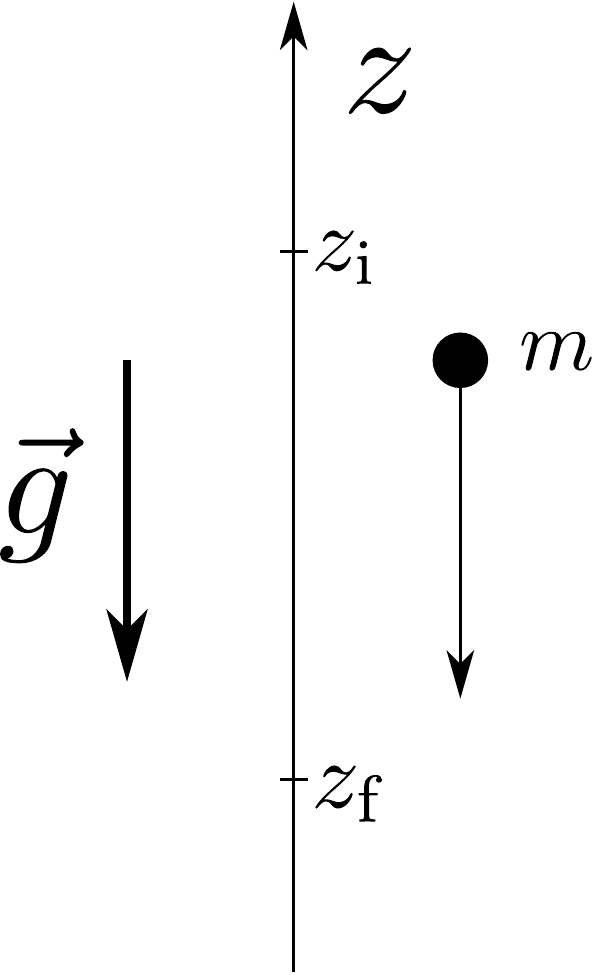}}}\hspace{50pt}
\subfigure[]{
\resizebox{!}{5.5cm}{\includegraphics{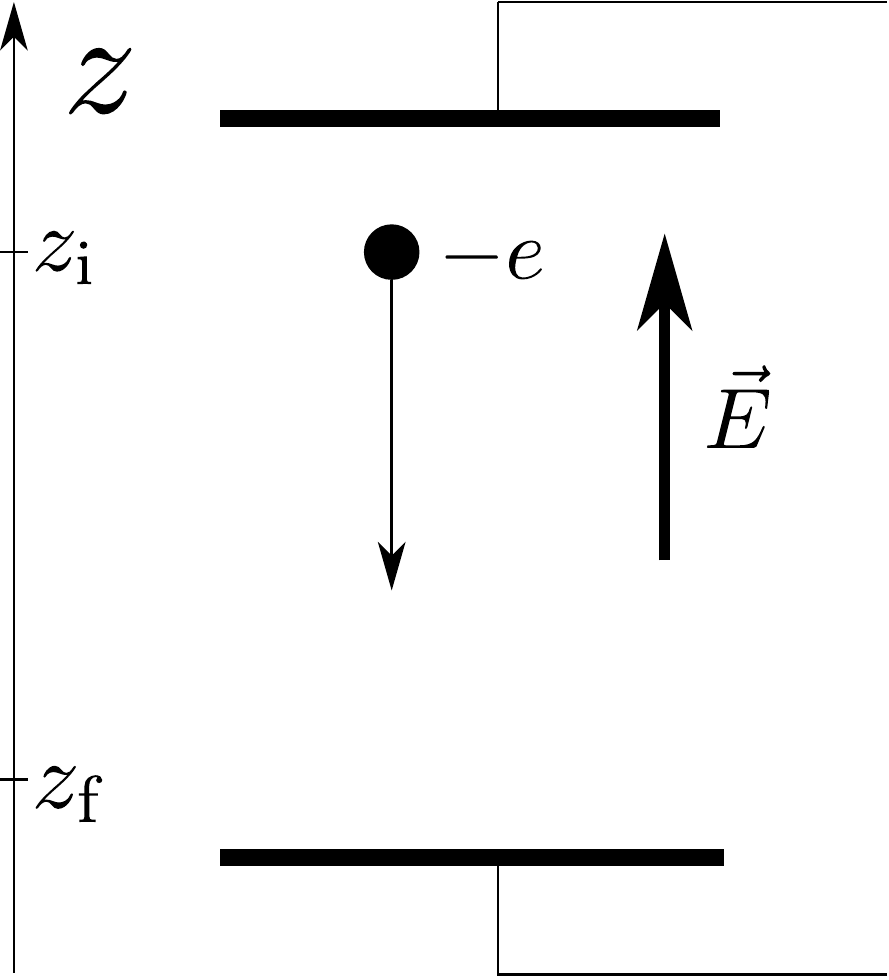}}}

\caption{Two physical systems with a linear potential $V(z)\equiv -Fz$ corresponding to a constant force ${\bf F}=F{\bf e}_z$ directed along the $z$-axis:
(a) a particle of mass $m$ which experiences the constant gravitational acceleration $g$ with $F\equiv -mg$,
and (b) a charge $-e$ in an ideal capacitor with the constant and homogeneous electric field $E$ with $F\equiv -eE$.}
\label{fig: g-and-electric}      
\end{figure*}
%

\subsection{Outline}
\label{subsec:1-3}

Our article is organized as follows. Section \ref{sec:2} serves as a motivation. Here we recall
that the propagator of a particle in a linear potential displays a phase which is cubic in time.
We show that this global phase depends on the initial wave function and outline an interferometric measurement scheme. 
In this arrangement the resulting interferometer phase which is cubic in time is independent of
the initial state of the center-of-mass motion.

In Section \ref{sec:3} we introduce our interferometer capable of measuring
the cubic phase accumulated by a particle during its motion in a linear potential provided by a constant gravitational field and magnetic field gradient.
The three-level atom probing these fields has two ground states corresponding to two different magnetic quantum numbers
and experiences a sequence of four Raman pulses.
In order to close the interferometer in position {\it and} velocity, we choose the separation of $T-2T-T$ between the pulses.
The resulting probability for the atom to exit the interferometer in one of the two ground states is the familiar oscillatory function,
which is then independent of the initial wave function.
However, in contrast to standard interferometers the argument now depends on the phase cubic in $T$ and the discrete third derivative of the laser phase.

We dedicate Section \ref{sec:4} to a comparison of the present scheme to the Kasevich-Chu interferometer
\cite{Kasevich-Chu-1,Kasevich-Chu-2,Giese,Schleich-Greenberger-Rasel,Gravity-inertial mass}
and distinguish the cubic phase shift from the ones caused by a gravity gradient or the Continuous-Acceleration Bloch (CAB) technique \cite{CAB technique}.
In Section \ref{sec:5} we discuss a possible experimental implementation of our proposal.
We conclude in Section \ref{sec:6} by briefly summarizing our results and providing an outlook.

In order to keep our article self-contained but focused on the central ideas, we include lengthy calculations in five appendices.
In Appendix \ref{sec:A} we recall the technique of creating coherent superpositions of the two ground states and interchanging
their populations using Raman pulses.
We then turn in Appendix \ref{sec:B} to a description of our interferometer as a sequence of unitary operators.
In Appendix \ref{sec:C} we derive the conditions to close our interferometer and obtain in Appendix \ref{sec:D} an explicit expression
for the interferometer phase induced by a linear potential.

Finally in Appendix \ref{sec:E} we make contact with the formalism of Ref. \cite{Roura2014} developed for a rather general form of interferometer
and rederive the phase of our $T^3$-interferometer. This approach brings out again that in contrast to the corresponding global phase
this phase is independent of the initial state of the center-of-mass motion.

\section{From global to interferometer phase}
\label{sec:2}

The propagator of a quantum particle experiencing a linear potential is determined by a phase factor governed by
the corresponding classical action. Since the relevant classical motion involves time in the coordinate and velocity
in a quadratic and a linear way, both the kinetic as well as potential energies bring in time quadratically.
As a result, the action being the integral over time must contain a term proportional to the cube of time. This cubic phase
which is independent of the coordinate is at the center of our interest in the present section.

We first recall the essential features of the propagator for the wave function in a linear potential. Here we focus especially
on this cubic phase. Moreover, we note that due to the Huygens principle for matter waves the integration over the initial coordinate
leads to a dependence of this phase on the initial wave function. 

Although we find this property interesting we emphasize that
it is of no importance for the present discussion. Indeed, due to the Born rule we cannot measure the global phase factor
of a {\it single} quantum system. However, an interferometric measurement of the difference of 
two different global phase factors of {\it two} systems is possible.
In such an interferometer the cubic phase is independent \cite{Roura2014} of the initial wave function.

\subsection{Emergence of $T^3$-phase in the propagator}
\label{subsec:2-1}

We start our analysis by discussing the propagator of a particle in a linear potential. Here we emphasize especially the emergence of the phase factor
cubic in time.

We consider a particle of mass $m$ moving in a linear potential $V(z)\equiv -Fz$ corresponding to a constant force ${\bf F}\equiv F{\bf e}_z$
directed along the $z$-axis with the unit vector ${\bf e}_z$. This problem occurs for
(i) a particle, which experiences a constant gravitational acceleration $g$ with $F\equiv -mg$ as indicated in Fig. \ref{fig: g-and-electric}a, and
(ii) a charge $-e$ in an ideal capacitor with the constant electric field $E$, for which $F\equiv -eE$ as shown in Fig. \ref{fig: g-and-electric}b.
Throughout this article we focus on the example of a linear gravitational potential.

The wave function
\begin{equation}
 \label{wave-function}
  \psi(z_\mathrm{f},t_\mathrm{f})=\int\limits_{-\infty}^{+\infty} G(z_\mathrm{f},t_\mathrm{f}|z_\mathrm{i},t_\mathrm{i})\psi(z_\mathrm{i},t_\mathrm{i})dz_\mathrm{i}
\end{equation}
representing the probability amplitude to find the particle
at the final position $z_\mathrm{f}$ at time $t_\mathrm{f}$ is determined by the propagator \cite{Kennard,Kennard2,Path-integral}
\begin{equation}
 \label{propagator}
  G(z_\mathrm{f},t_\mathrm{f}|z_\mathrm{i},t_\mathrm{i})\equiv \langle z_\mathrm{f}|\exp\left[-\frac{i}{\hbar}\left(\frac{\hat{p}_z^2}{2m}-F\hat{z}\right)(t_\mathrm{f}-t_\mathrm{i})\right]|z_\mathrm{i}\rangle,
\end{equation}
where $\psi(z_\mathrm{i},t_\mathrm{i})$ is the value of the wave function at the initial position $z_\mathrm{i}$ and time $t_\mathrm{i}$, and
$\hat{z}$ and $\hat{p}_z$ denote the position and momentum operators, respectively.

The propagator $G$ defined by Eq. (\ref{propagator}) can be cast \cite{Path-integral}
in terms of the classical action
\begin{equation}
 \label{Scl}
  \begin{split}&S_{\mathrm{cl}}(z_\mathrm{f},t_\mathrm{f}|z_\mathrm{i},t_\mathrm{i})\equiv\int\limits_{t_\mathrm{i}}^{t_\mathrm{f}}{L}\left(z_\mathrm{cl}(t),{\dot z}_\mathrm{cl}(t)\right) dt\\
  &=\frac{m}{2}\frac{(z_\mathrm{f}-z_\mathrm{i})^2}{t_\mathrm{f}-t_\mathrm{i}}+\frac{F}{2}\left(z_\mathrm{f}+z_\mathrm{i}\right)(t_\mathrm{f}-t_\mathrm{i})-\frac{F^2}{24 m}(t_\mathrm{f}-t_\mathrm{i})^3\end{split}
\end{equation}
along the classical trajectory given by
\begin{equation}
 \label{cl-trajectory-z}
  z_{\mathrm{cl}}(t)\equiv z_\mathrm{i}+\frac{z_\mathrm{f}-z_\mathrm{i}}{t_\mathrm{f}-t_\mathrm{i}}(t-t_\mathrm{i})+\frac{F}{2m}(t-t_\mathrm{i})(t-t_\mathrm{f})
\end{equation}
and
\begin{equation}
 \label{cl-trajectory-v}
  {\dot z}_\mathrm{cl}(t)\equiv \frac{d}{dt}z_\mathrm{cl}(t)=\frac{z_\mathrm{f}-z_\mathrm{i}}{t_\mathrm{f}-t_\mathrm{i}}+\frac{F}{m}\left(t-\frac{t_\mathrm{i}+t_\mathrm{f}}{2}\right).
\end{equation}
Here we have used the Lagrange function
\begin{equation}
 \label{Lagrange function}
  L\left(z,{\dot z}\right)\equiv \frac{m}{2}{\dot z}^2+Fz
\end{equation}
of a particle in a linear potential.

Indeed, the representation
\begin{equation}
 \label{propagator-cl}
  G(z_\mathrm{f},t_\mathrm{f}|z_\mathrm{i},t_\mathrm{i})=N(t_\mathrm{f}-t_\mathrm{i})
  \exp\left[\frac{i}{\hbar}S_{\mathrm{cl}}(z_\mathrm{f},t_\mathrm{f}|z_\mathrm{i},t_\mathrm{i})\right]
\end{equation}
with the normalization
\begin{equation}
 \label{propagator-cl-normalization}
  N(t)\equiv\sqrt{\frac{m}{2i\pi \hbar t}}
\end{equation}
brings out most clearly that $G$ contains the phase
\begin{equation}
 \label{phase}
  \phi\equiv -\frac{1}{24}\frac{F^2}{\hbar m}\,t^3,
\end{equation}
which is independent of the initial and final positions $z_\mathrm{i}$ and $z_\mathrm{f}$, and scales with
the third power of the time difference $t\equiv t_\mathrm{f}-t_\mathrm{i}$,
that is the time during which the particle experiences the constant force $F$.


\begin{figure*}
\centering
\resizebox{0.5\textwidth}{!}{\includegraphics{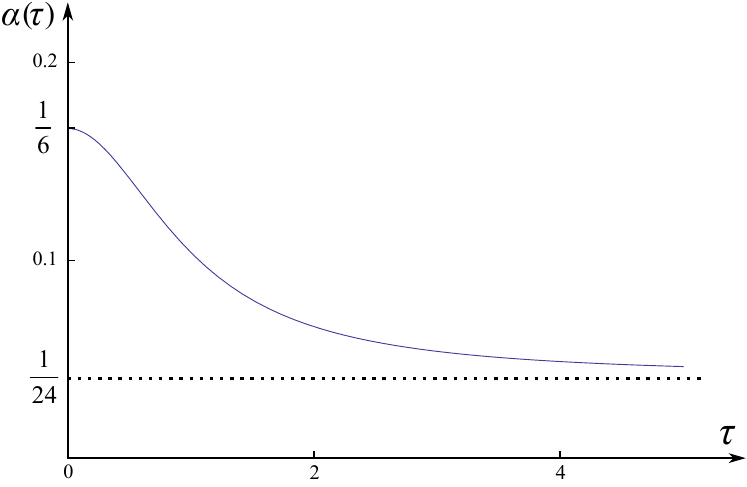}}

\caption{Dependence of the numerical factor $\alpha=\alpha(\tau)$ defined by Eq. (\ref{coefficient alpha}) on the dimensionless ratio $\tau$
of coordinate time $t$ and spreading time $t_s$ given by Eq. (\ref{t spreading}). For $\tau\rightarrow 0$ and $\tau\rightarrow \infty$ the factor $\alpha$
is almost constant and given by $1/6$ and $1/24$ (dashed line), respectively.
However, for values of $\tau$ between these extremes $\alpha$ changes rapidly and thus in this transition domain the phase
$\tilde{\phi}=\tilde{\phi}(t)$, Eq. (\ref{wave function phase result}), is not strictly cubic.}
\label{fig: alpha}       
\end{figure*}
%


\subsection{Dependence of $T^3$-phase on initial wave function}
\label{subsec:2-2}

The cubic phase $\phi$ in the propagator of a quantum particle moving in a linear potential
manifests itself in every wave function exposed to this situation. Indeed, since $\phi$ is independent of
the initial coordinate $z_\mathrm{i}$ it can be factored out of the Huygens integral for matter waves, Eq. (\ref{wave-function}).

Nevertheless, the integration of the remaining parts of the propagator in combination with the initial wave function can change
the numerical factor in the cubic phase.
In order to address this feature in more detail we consider the normalized initial wave function
\begin{equation}
 \label{wave function initial}
  \psi(z_\mathrm{i},t_\mathrm{i})\equiv \frac{1}{\left(\sqrt{\pi}\Delta z_0\right)^{1/2}}\exp\left(-\frac{z_\mathrm{i}^2}{2\Delta z_0^2}\right)
\end{equation}
in the form of a Gaussian of width $\Delta z_0$.

When we substitute this expression into the Huygens integral of matter waves, Eq. (\ref{wave-function}), and
use the expressions, Eqs. (\ref{Scl}) and (\ref{propagator-cl}), for the propagator we arrive at the final wave function
\begin{equation}
 \label{wave function result}
  \psi(z_\mathrm{f},t_\mathrm{f})=\frac{1}{\left[\sqrt{\pi}\Delta z(t)\right]^{1/2}}
  \exp\left\{-\frac{\left(z_\mathrm{f}-\frac{F}{2m}t^2\right)^2}{2\left[\Delta z(t)\right]^2}+i\beta\right\}
\end{equation}
with the time-dependent width
\begin{equation}
 \label{wave function width}
  \Delta z(t)\equiv\Delta z_0\sqrt{1+\frac{t^2}{t_{\rm s}^2}}
\end{equation}
and phase
\begin{equation}
 \label{wave function phase}
  \beta(t)\equiv \frac{F z_\mathrm{f} t}{\hbar}+
  \frac{t}{t_{\rm s}}\frac{\left(z_\mathrm{f}-\frac{F}{2m}t^2\right)^2}{2\left[\Delta z(t)\right]^2}-
  \frac{F^2t^3}{6\hbar m}-\frac{1}{2}\arctan\left(\frac{t}{t_{\rm s}}\right).
\end{equation}
Here
\begin{equation}
 \label{t spreading}
  t_{\rm s}\equiv \frac{m\Delta z_0^2}{\hbar}
\end{equation}
denotes the spreading time of the wave packet.

We combine the term proportional to $t^5$ arising from the square in the second contribution to Eq. (\ref{wave function phase})
with the cubic phase and find the total phase
\begin{equation}
 \label{wave function phase result}
    \tilde{\phi}(t)\equiv -\alpha\left(\frac{t}{t_s}\right)\frac{F^2}{\hbar m}\,t^3.
\end{equation}
Here we have introduced the time-dependent numerical factor
\begin{equation}
 \label{coefficient alpha}
    \alpha\left(\tau\right)\equiv \frac{1}{24} \frac{\tau^2+4}{\tau^2+1}
\end{equation}
depending on the dimensionless ratio $\tau$ of coordinate time and spreading time $t_s$,
which according to Eq. (\ref{t spreading}) is proportional to square of the initial width $\Delta z_0$.

For a plane wave we find $\Delta z_0\rightarrow \infty$ and thus $t_s\rightarrow \infty$ leading us to
\begin{equation}
 \label{coefficient alpha plane wave}
    \alpha\left(\tau\rightarrow 0\right)= \frac{1}{6}\,.
\end{equation}

However, for an infinitely narrow Gaussian with $\Delta z_0\rightarrow 0$ and thus $t_s\rightarrow 0$ we find
\begin{equation}
 \label{coefficient alpha small}
    \alpha\left(\tau\rightarrow \infty\right)= \frac{1}{24}\,.
\end{equation}

So far we have only considered the extreme cases of $\tau$.
Only in the domains where $\alpha$ is approximately constant do we find a pure cubic phase dependence on $t$.
Indeed, between the extremes the time dependence is more complicated as expressed in Fig. 2.

\subsection{How to observe the $T^3$-phase?}
\label{subsec:2-3}

The propagator $G$ defined in Eq. (\ref{propagator-cl}) contains a global phase which is cubic in time. Due to the integration
over the initial position in the Huygens integral, Eq. (\ref{wave-function}), this phase depends on the initial wave function.
These two sentences summarize in a pregnant way our results of the preceding section. We now briefly outline our interferometric
measurement strategy for this phase. In the course of our analysis we shall find that the phase in this interferometer
is independent of the initial state.

Obviously a setup providing us only with the probability density $|\psi(z_\mathrm{f},t_\mathrm{f})|^2$ is insensitive
to any global phase like the $T^3$-phase. Therefore, we need to involve an interferometric measurement
either with a path-dependent mass of the particle, or a path-dependent strength of the constant force.

Throughout this article we focus on the second alternative although we can imagine possibilities to utilize
the dependence of the mass on the internal state. Key elements of our technique are:
(i) an atom with both a magnetically sensitive and a magnetically insensitive internal states and
(ii) an external time-independent magnetic field with a constant gradient along one direction.
Due to the Zeeman effect, such an atom experiences a constant force determined by its internal state.
It is the same force that acts on a classical magnetic dipole in a non-uniform magnetic field.

\section{Atom interferometer with four light pulses}
\label{sec:3}

In this section we introduce the two crucial elements of our $T^3$-interferometer:
(i) the Zeeman shift of the atomic levels induced by the external time-independent magnetic field with a constant gradient along one direction,
and (ii) the population dynamics of the two resonant states of the atom driven by the Raman laser pulses.

\subsection{Zeeman effect: control of external degrees of freedom}

We consider a three-level atom consisting of the ground state $|g_1\rangle$, the state $|g_2\rangle$, and the excited state $|e\rangle$,
as indicated in Fig. \ref{fig: scheme}. Here $|g_1\rangle$ and $|g_2\rangle$ are chosen such that
the mean value of their magnetic moment differs, which leads to a state-dependent acceleration.
Without loss of generality, we simplify the following calculations by selecting a magnetic insensitive state for $|g_1\rangle$.
The center-of-mass motion of the atom is assumed to be along the $z$-axis, which is the direction of the constant gravitational acceleration.

The interaction of the atom with a time-independent magnetic field having locally the form\footnote[2]{Throughout the article,
we use the notation $\nabla_z B_z\equiv \frac{\partial B_z}{\partial z}({\bf r}=0)$
for the derivative of the $z$-component of the magnetic field ${\bf B}={\bf B}({\bf r})$ along the $z$-direction at the origin ${\bf r}=0$.
This derivative is assumed to be small compared to $B_0$, such that $L|\nabla_z B_z|\ll |B_0|$,
where $L$ is the total length of the interferometer.
Moreover, we note that the form  of the magnetic field given by Eq. (\ref{B-field}) is an approximate one.
Indeed, according to the Maxwell equation $\nabla \cdot{\bf B}=0$, which is valid everywhere,
a non-zero value of $\nabla_z B_z$ induces non-zero values of $\nabla_x B_x$ and $\nabla_y B_y$,
such that $\nabla_x B_x+\nabla_y B_y=-\nabla_z B_z$, where $B_x$ and $B_y$ are the components of ${\bf B}$ along the $x$- and $y$-axis.
However, in the limit of $L|\nabla_z B_z|\ll|B_0|$
the magnetic field ${\bf B}$ given by Eq. (\ref{B-field}) is approximately directed along the $z$-axis.}
\begin{equation}
 \label{B-field}
  {\bf B}({\bf r})\cong\left(B_0+z\,\nabla_z B_z\right){\bf e}_z
\end{equation}
results in the linear Zeeman shift
\begin{equation}
 \label{Zeeman-shift1}
  \Delta E_{g_2}^{Z}=\mu_B g_{L} m_{g_2}\left(B_0+z\,\nabla_z B_z\right)
\end{equation}
of the energy of $|g_{2}\rangle$, and no Zeeman shift
\begin{equation}
 \label{Zeeman-shift2}
  \Delta E_{g_1}^Z=0
\end{equation}
of the energy of $|g_{1}\rangle$. Here $\mu_B$, $g_{L}$, $m_{g_1}$, and $m_{g_2}$ denote the Bohr magneton,
the Land$\rm\acute{e}$ $g$-factor of $|g_2\rangle$,
and the magnetic quantum numbers associated with the $z$-component of the angular momentum
for $|g_1\rangle$ and $|g_2\rangle$, which are given by $m_{g_1}=0$ and $|m_{g_2}|>0$, respectively.

When we take into account the linear gravitational potential $V(z)=mgz$ we arrive at the Hamiltonian
\begin{equation}
 \label{Hamiltonian-free atom}
 \begin{split}& {\hat H}_{\mathrm{at}}\equiv \mathds{1}_3\otimes\frac{{\hat p}_z^2}{2m}+\left[\left(E_e^{(0)}-m a_1 \hat{z}\right)|e\rangle\langle e| \right.\\
&\left.+\left(E_{g_1}^{(0)}-m a_1 \hat{z}\right)|g_1\rangle \langle g_1|\right.\\
&\left.+\left( E_{g_2}^{(0)}+\hbar\omega_0-m a_2 \hat{z}\right)|g_2\rangle \langle g_2|\right] \otimes\mathds{1}_z,\end{split}
\end{equation}
of the three-level atom, where
$$
\mathds{1}_3\equiv |e\rangle\langle e|+|g_1\rangle\langle g_1|+|g_2\rangle\langle g_2|$$
and
$$
\mathds{1}_z\equiv\int_{-\infty}^{+\infty} dz|z\rangle\langle z|
$$
are the identity operators corresponding to the Hilbert space of the internal atomic states, and the center-of-mass motion along the $z$-axis, respectively.
Here $E_{e}^{(0)}$, $E_{g_1}^{(0)}$, and $E_{g_2}^{(0)}$ are the energies of the atom
in the internal states $|e\rangle$, $|g_1\rangle$, and $|g_2\rangle$, respectively, when the atom is not exposed to any external field.

The homogeneous magnetic field $B_0$ leads to an energy shift $\hbar\,\omega_0\equiv\mu_B g_L m_{g_2}B_0$ and
\begin{equation}
 \label{a1-a2}
  a_1\equiv -g\;\;\;\;\;{\rm and}\;\;\;\;\;a_2\equiv -g-\frac{\mu_B }{m} g_L m_{g_2}\nabla_z B_z
\end{equation}
denote the accelerations of the atomic center-of-mass corresponding to $|g_1\rangle$ and $|g_2\rangle$, respectively.
Thus, the Zeeman effect for $|g_1\rangle$ and $|g_2\rangle$ in the magnetic field of constant gradient gives rise to the two different accelerations $a_{1}$ and $a_{2}$.

\subsection{Population dynamics}

\begin{figure*}
\centering
\resizebox{0.9\textwidth}{!}{\includegraphics{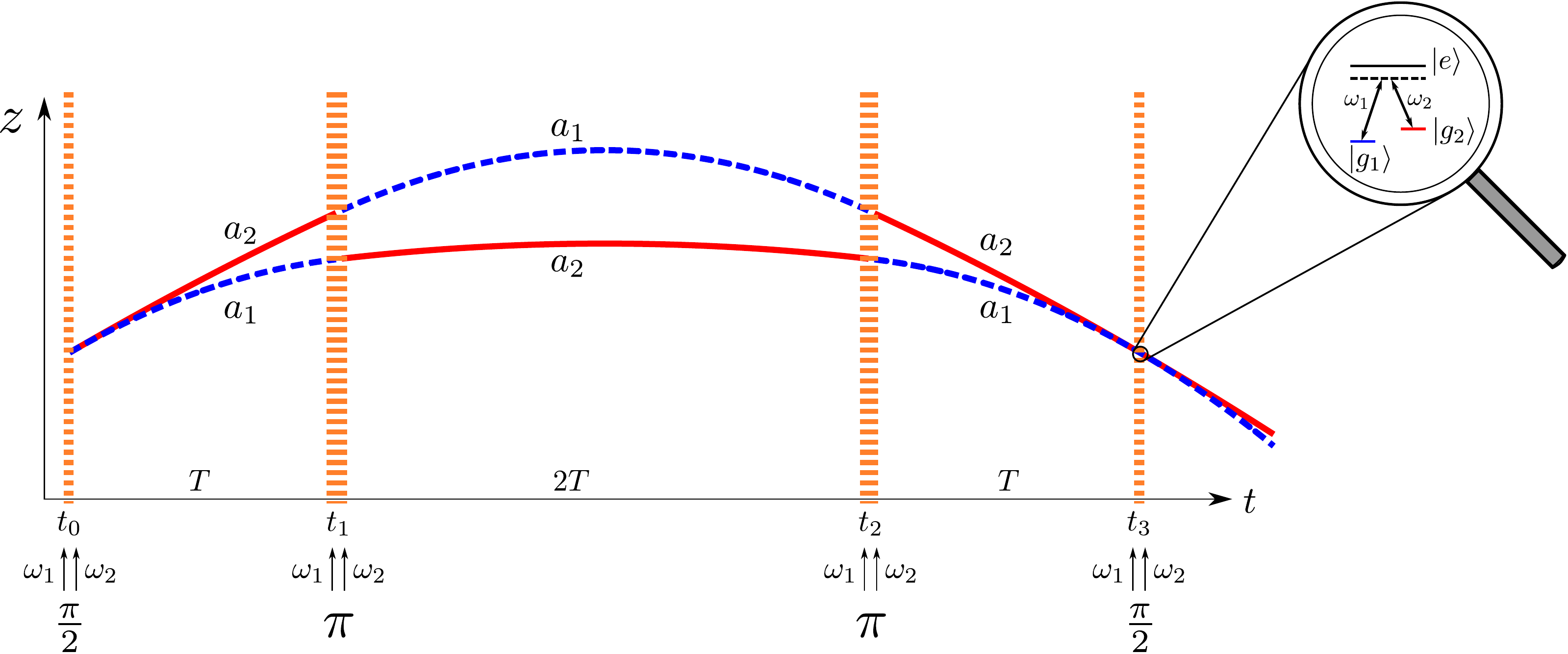}}

\caption{Space-time diagram of the $T^3$-interferometer for a three-level atom consisting of the states
$|g_1\rangle$, $|g_2\rangle$, and $|e\rangle$, and interacting with four short Raman laser pulses
at $t=t_0$, $t=t_1\equiv t_0+T$, $t=t_2\equiv t_0+3T$ and $t=t_3\equiv t_0+4T$.}
\label{fig: scheme}       
\end{figure*}

We are now in the position to present our $T^3$-interferometer for atoms capable of measuring the cubic phase.
The general scheme of this particular atom interferometer depicted in Fig. \ref{fig: scheme} consists of two distinct ``building blocks'':
(i) four Raman pulses, that is two $\frac{\pi}{2}$- and two $\pi$-pulses, which form a $\frac{\pi}{2}-\pi-\pi-\frac{\pi}{2}$ sequence, and
(ii) three regions of the atomic center-of-mass motion with constant accelerations $a_1$ and $a_2$.

Each block is described in terms of an appropriate unitary operator. As discussed in more detail in Appendix \ref{sec:A},
the atom-light interaction is accounted for by the interaction operator
\begin{equation}
 \label{Atom-light-evaluation}
 \begin{split}& \hat{U}_\mathrm{p}(\theta)\equiv\left(\,|g_1\rangle\langle g_1|+|g_2\rangle\langle g_2|\,\right)\cos\left(\frac{\theta}{2}\right)\\
 &-i\left(e^{i\phi_\mathrm{L}}|g_1\rangle\langle g_2|+e^{-i\phi_\mathrm{L}}|g_2\rangle\langle g_1|\right)\sin\left(\frac{\theta}{2}\right), \end{split}
\end{equation}
where $\phi_\mathrm{L}$ is the laser phase and $\theta$ denotes the total pulse area. Moreover, we have assumed that
each Raman pulse consists of two co-propagating laser beams with nearly equal wavelengths which makes
$\hat{U}_{\mathrm{p}}$ given by Eq. \eqref{Atom-light-evaluation} independent of $z$.

On the other hand, the operator
\begin{equation}
 \label{operator-U-gravity}
  \hat{U}_a(t_\mathrm{f},t_\mathrm{i})\equiv \exp\left[-\frac{i}{\hbar}\left(\frac{\hat{p}_z^2}{2m}-ma\hat{z}\right)(t_\mathrm{f}-t_\mathrm{i})\right]
\end{equation}
with $a=a_1$ or $a=a_2$ given by Eq. (\ref{a1-a2}) provides us with the center-of-mass motion.


%


In Appendices \ref{sec:B} and \ref{sec:C} we analyse the interferometer of Fig. \ref{fig: scheme} as a sequence of these unitary operators and find the expression
\begin{equation}
 \label{interferometer-probability-result}
  P_{g_2}=\frac{1}{2}\left[1+\cos\left(\varphi_\mathrm{i}+\varphi_\mathrm{L}\right)\right]
\end{equation}
for the probability of observing the atoms in the excited state $|g_2\rangle$ after the action of the four Raman pulses.
According to Appendix \ref{sec:D} the interferometer phase $\varphi_\mathrm{i}$ reads
\begin{equation}
 \label{interferometer-probability-interferometer phase}
  \varphi_\mathrm{i}\equiv\frac{m}{\hbar}(a_1^2-a_2^2)T^3,
\end{equation}
and in Appendix \ref{sec:B} we derive the expression
\begin{equation}
 \label{interferometer-probability-laser-phase}
  \varphi_\mathrm{L}\equiv \phi_\mathrm{L}(t_0)-2\phi_\mathrm{L}(t_0+T)+2\phi_\mathrm{L}(t_0+3T)-\phi_\mathrm{L}(t_0+4T)
\end{equation}
for the total laser phase.

We emphasize that this result is independent of the initial state of the center-of-mass motion.
This property of the {\it interferometer phase} is in sharp contrast to the dependence of global phase of Sec. \ref{subsec:2-2} corresponding
to a {\it single} degree of freedom. As shown in Ref. \cite{Roura2014} it is a consequence of the fact that
our interferometer is closed in both position {\it and} velocity. Indeed, in Appendix \ref{sec:E} we use the expression
for the interferometer phase of a rather general interferometer obtained in Ref. \cite{Roura2014}
to rederive Eq. (\ref{interferometer-probability-interferometer phase}).

We conclude by recalling Eq. \eqref{a1-a2} to obtain the explicit dependence of $\varphi_\mathrm{i}$ on the magnetic field gradient and the gravitational acceleration and cast Eq. \eqref{interferometer-probability-interferometer phase} into the form
\begin{equation}
\label{interferometer-probability-interferometer phase-g}
\varphi_\mathrm{i}=-\frac{\mu_B}{\hbar}g_L m_{g_2}\nabla_z B_z\left(2g+\frac{\mu_B}{m}g_L m_{g_2}\nabla_z B_z\right)T^3\;.
\end{equation}
For a known magnetic field gradient the gravitational acceleration can then be measured with a cubic scaling in $T$.
By using different Zeeman sub-levels for the state  $|g_2\rangle$ it is even possible to measure both, the magnetic field gradient $\nabla_z B_z$ {\it and} the gravitational acceleration $g$. A simultaneous measurement of these quantities has already been demonstrated for a Bragg interferometer with an interferometer phase scaling as $T^2$ \cite{Hardman}.

\section{Discussion}
\label{sec:4}

In the preceding section we have derived an expression for the probability of finding the atom in the state $|g_2\rangle$
at one exit of our interferometer. In the present section we compare and contrast our device with the Kasevich-Chu interferometer,
and make contact with other cubic phases such as the ones caused by a gravity gradient, or arising in the CAB technique.

\subsection{Comparison with Kasevich-Chu interferometer}
\label{subsec:3-1}

In the Kasevich-Chu interferometer the probability corresponding to Eq. \eqref{interferometer-probability-result}
reads \cite{Kasevich-Chu-1,Kasevich-Chu-2,Giese,Schleich-Greenberger-Rasel}
\begin{equation}
 \label{interferometer-probability-result-KC}
  P_{g_2}^{(KC)}=\frac{1}{2}\left[1-\cos\left(\varphi_\mathrm{i}^{(KC)}+\varphi_\mathrm{L}^{(KC)}\right)\right]
\end{equation}
with the interferometer phase
\begin{equation}
 \label{interferometer-probability-interferometer phase-KC}
  \varphi_\mathrm{i}^{(KC)}\equiv (k_1+k_2)gT^2
\end{equation}
and the total laser phase
\begin{equation}
 \label{interferometer-probability-laser-phase-KC}
  \varphi_\mathrm{L}^{(KC)}\equiv \phi_\mathrm{L}(t_0)-2\phi_\mathrm{L}(t_0+T)+\phi_\mathrm{L}(t_0+2T).
\end{equation}

We note four major differences between our scheme and that of Kasevich-Chu:
(i) The {\it four} rather than the {\it three} Raman pulses create the \textit{sum} of the two terms appearing in the square brackets of $P_{g_2}$
given by Eq. (\ref{interferometer-probability-result}) rather than the {\it difference} in $P_{g_2}^{(KC)}$ defined by Eq. (\ref{interferometer-probability-result-KC}).
In the absence of any potentials the different pulse sequences can be visualized by a $2\pi$-rotation on the Bloch sphere
in the case of the Kasevich-Chu interferometer, and a $3\pi$-rotation for our $T^3$-interferometer giving rise to the opposite signs.
(ii) The phase $\varphi_\mathrm{i}$ following from Eq. (\ref{interferometer-probability-interferometer phase}) and induced by the linear potentials
depends on the separation $T$ of the pulses in a {\it cubic} rather than {\it quadratic} way as in $\varphi_\mathrm{i}^{(KC)}$
expressed by Eq. \eqref{interferometer-probability-interferometer phase-KC}.
(iii) {\it Co-propagating} laser beams together with a constant magnetic field gradient lead to a proportionality of $\varphi_\mathrm{i}$ to $\nabla_zB_z$,
while in the case of Kasevich-Chu the use of {\it counter-propagating} laser beams results in a momentum transfer of $\pm \hbar(k_1+k_2)$
which reflects itself in $\varphi_\mathrm{i}^{(KC)}$. (iv) The total laser phase $ \varphi_\mathrm{L}$ defined by Eq. (\ref{interferometer-probability-laser-phase})
is a discrete {\it third} derivative rather than the {\it second} one for $\varphi_\mathrm{L}^{(KC)}$ given by Eq. (\ref{interferometer-probability-laser-phase-KC}).
Indeed, this feature becomes obvious when we consider the limit of  $T\rightarrow 0$, for which $\varphi_\mathrm{L}\cong -2{\dddot\varphi}_\mathrm{L}(t_0) T^3$ while \cite{Giese,Schleich-Greenberger-Rasel}
$\varphi_\mathrm{L}^{(KC)}\cong {\ddot\varphi}_\mathrm{L}(t_0) T^2$, with
${\dddot\varphi}_\mathrm{L}(t_0)$ and ${\ddot\varphi}_\mathrm{L}(t_0)$ being the third and second continuous derivatives
of the phase $\varphi_{\mathrm{L}}=\varphi_\mathrm{L}(t)$ of the Raman pulse, respectively.

\subsection{Other origins of cubic phases}
\label{subsec:3-2}

We now compare and contrast the $T^3$-phase in our interferometer induced by the propagator of a particle in a linear potential
to other phases cubic in time. Here we focus on two different situations:
(i) the presence of a gravity gradient, or (ii) the application of the CAB technique.

\subsubsection{Gravity gradient}
\label{subsubsec:3-2-1}

In the presence of a gravity gradient $\Gamma$ a phase cubic in time appears in the Kasevich-Chu interferometer as a consequence of a {\it single quadratic} potential
\begin{eqnarray}
  V_\Gamma(z)\equiv mgz+\frac{1}{2}m\Gamma z^2
\end{eqnarray}
rather than {\it two linear} potentials. In particular, it results from an expansion of the atomic center-of-mass motion
in the limit of a weak gravity gradient, that is $\Gamma T^2\ll 1$.
Indeed, a gravity gradient leads to a {\it position-dependent} acceleration,
while the two linear potentials in our interferometer lead to a {\it state-dependent} acceleration.

Furthermore, we emphasize, that the Kasevich-Chu interferometer is no longer closed in position {\it and} velocity in the presence of a gravity gradient.
This deficiency leads to a loss of contrast \cite{Roura2014} and a dependence of the phase shift on the initial state.

The interferometer can be closed (at least approximately) by additional laser pulses \cite{Marzlin1996} or by suitably adjusting the laser wavelength
of the intermediate pulse \cite{Roura2015}. However, this procedure also eliminates the cubic contributions to the phase shift,
in contrast to the situation considered here.


\begin{figure*}
\centering
\resizebox{0.7\textwidth}{!}{%
  \includegraphics{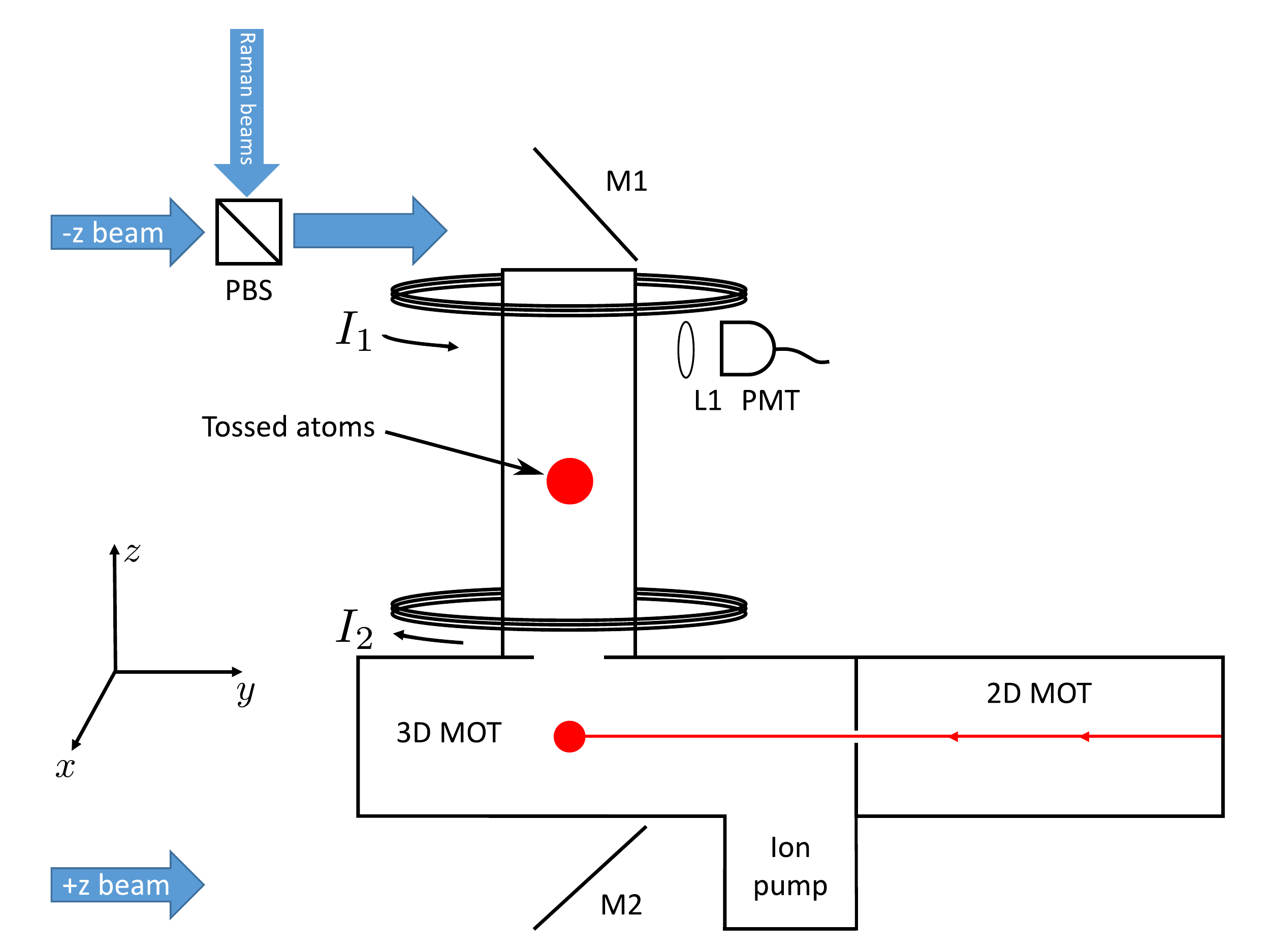}
}

\caption{Experimental setup for the $T^3$-interferometer. Our $^{85}{\rm Rb}$ atoms emerge from a two-dimensional magneto-optical trap (2D MOT),
pass through an aperture into the 3D MOT as indicated by the red line, and are then launched into a glass tower in a moving molasses configuration.
The mirrors $M_1$ and $M_2$ reflect the $+z$ and $-z$ beams of the 3D MOT, respectively, as well as the two co-propagating Raman beams
with the parallel circular polarizations, used for the control of the internal atomic states.
We employ two coils in an anti-Helmholtz configuration with currents $I_1$ and $I_2$ to create a magnetic field
with constant gradient in the $z$-direction along the glass tower.
A solenoid not depicted here surrounds the tower to provide an additional nominally uniform magnetic field.
The lens $L_1$ collects the fluorescence from the tossed atoms at the top of the tower and focuses this light onto a photomultiplier (PMT).}
\label{fig: experimental setup}       
\end{figure*}
%


\subsubsection{Continuous-Acceleration-Bloch technique}
\label{subsubsec:3-2-2}

Our $T^3$-interferometer shares the underlying idea of CAB, that is applying different constant accelerations along each interferometer arm.
However, instead of achieving these accelerations via state-dependent linear potentials,
a beam splitter based on Bragg-diffraction is used to load one of the two exit ports into an optical lattice,
 which is accelerated subsequently by the use of Bloch oscillations.

While in our scheme we can close the interferometer easily in position {\it and} velocity by simply choosing the correct timing between the pulses, the CAB scheme requires a sophisticated control of the acceleration of the optical lattice. Moreover, we emphasize that in contrast to the $T^3$-interferometer, in the CAB scheme not only a phase proportional to $T^3$, but also one proportional to $T^2$ emerges.

\section{Towards an experimental realization}
\label{sec:5}

In this section we discuss a possible experimental implementation of our proposal for a $T^3$-interferometer
based on our current laboratory apparatus depicted in Fig. \ref{fig: experimental setup}.
We first summarize the key features of our setup, describe our method to deduce the magnetic fields from the observed Raman spectra,
and conclude by briefly analysing the present limitation of our device due to decoherence.

\subsection{Experimental setup}
\label{subsec:4-1}

We use the D2 transition of $^{85}{\rm Rb}$ and choose $|g_{1}\rangle$ and $|g_{2}\rangle$ from the $F=2$ and $F=3$ hyperfine state manifolds
with a frequency separation of approximately $3{\rm\, GHz}$ \cite{Frank-JMO2013}.
The atoms are loaded into a three-dimensional magneto-optical trap (3D MOT) emerging from a two-dimensional trap (2D MOT)
as shown in Fig. \ref{fig: experimental setup}. The 3D MOT consists of standard cooling and repump beams as well as magnetic coils.
We rely on an all-glass chamber as our vacuum system.

After  $1 {\rm\, s}$ of loading to obtain a sufficient signal-to-noise ratio, the atoms are launched upwards along the $z$-axis
in a moving optical molasses configuration with a velocity of approximately $3 {\rm\, m/s}$ such that they strike the top of a $10 {\rm\, cm}$ tall glass tower.
It takes the atoms between $20 {\rm\, ms}$ and $40 {\rm\, ms}$ to reach this point depending on the launch velocity
which can be adjusted by the voltage of the launch signal.
We emphasize that the top of the tower does not coincide with the apex of the trajectory.

After launch the atoms are freely moving in the dark and we are able to apply a single or several Raman pulses 
involving two co-propagating laser beams along the $z$-axis with the same circular polarization. 
During their motion the atoms interact with a magnetic field which varies linearly along the $z$-axis due to coils in an anti-Helmholtz configuration
which are not the ones used to trap the atoms.
Moreover, they feel the field of a solenoid of finite length not depicted in Fig. \ref{fig: experimental setup} surrounding the glass tower.
The ability to change the current independently in each of the gradient coils and the solenoid provides us
with the control of the location of the zero crossing of the magnetic field, or an effective way to adjust the bias field.

On the top of the tower a photomultiplier tube (PMT) detector performs a projective measurement of the population
in the state $|g_{2}\rangle$ by collecting the fluorescence emitted by the atoms
caused by the vertical trapping beams, that is the $+z$ and $-z$ beams, which are switched back on for the measurement.
Observing the atomic population always at the top of the tower provides us with a convenient way of varying
the position where the Raman pulse is applied to the atoms along their flight path.

\subsection{Raman spectrum}
\label{subsec:4-2}

With this setup we can map out the magnetic field along the atom trajectory by measuring Raman spectra of the type
shown in Fig. \ref{fig: Raman spectrum}. For this purpose we first use an optical pumping stage to transfer all atoms
to the ground state $|g_{1}\rangle$, that is the state with $F=2$.
Then we apply a Raman light pulse whose intensity and duration are chosen to be close to a zero-detuning $\pi$-pulse.
We have found that this condition is satisfied for a pulse duration of $25-100 \,\mu{\rm s}$,
using our typical total Raman power of approximately $80\, {\rm mW}$
in a beam of diameter $2.5\, {\rm cm}$ and a single photon detuning of $1-2\, {\rm GHz}$.
Finally, we observe the number of atoms transferred from $|g_{1}\rangle$ to $|g_{2}\rangle$, that is the state with $F=3$.

In this manner we obtain Raman spectra such as the one presented in Fig. \ref{fig: Raman spectrum},
that is the population of $|g_2\rangle$ versus the two-photon detuning
\begin{equation}
 \label{Two-photon-detuning}
   \tilde{\Delta}\equiv \frac{E_{3,0}-E_{2,0}}{\hbar}+\omega_{2}-\omega_{1}
\end{equation}
of the Raman pulse of the frequencies $\omega_1$ and $\omega_2$ with respect to
the ``clock'' transition $|F_1=2,m_{F_1}=0\rangle\to|F_2=3,m_{F_2}=0\rangle$, for which the resonance occurs at $\tilde{\Delta}=0$.
Here $E_{F,m_{F}}$ denotes the energy of the hyperfine state.

For a magnetic field pointing in an arbitrary direction, the observed Raman spectrum displays up to 11 peaks;
that is 5 peaks for $\Delta m_F=0$ transitions and 6 peaks for $\Delta m_F=\pm 1$ transitions,
where $\Delta m_F\equiv m_{F,2}-m_{F,1}$ is the change of the magnetic quantum number.
For a magnetic field directed along the $z$-axis as required for the $T^3$-experiment,
transitions with $\Delta m_F=\pm 1$ are suppressed in our experimental setup \cite{Frank-JMO2013,Frank-JMO2011}.

\subsection{Magnetic field measurement}
\label{subsec:4-3}

The position of the relative two-photon resonance in the Raman spectrum
corresponding to the magnetically sensitive transition $|F_1,m_{F_1}\rangle\to|F_2,m_{F_2}\rangle$
is determined by the first-order Zeeman shift
\begin{equation}
 \label{detuning shift}
    \tilde{\Delta}_{F_1,m_{F_1}}^{F_2,m_{F_2}}\equiv\frac{\mu_B}{\hbar}\left(g_{F_2}m_{F_2}-g_{F_1}m_{F_1}\right)B(z)
\end{equation}
in frequency, where $B(z)$ is the value of the magnetic field at the center-of-mass coordinate $z$ of the atom cloud
during the interaction with the Raman pulse.

\begin{figure*}
\centering
\resizebox{0.9\textwidth}{!}{%
  \includegraphics{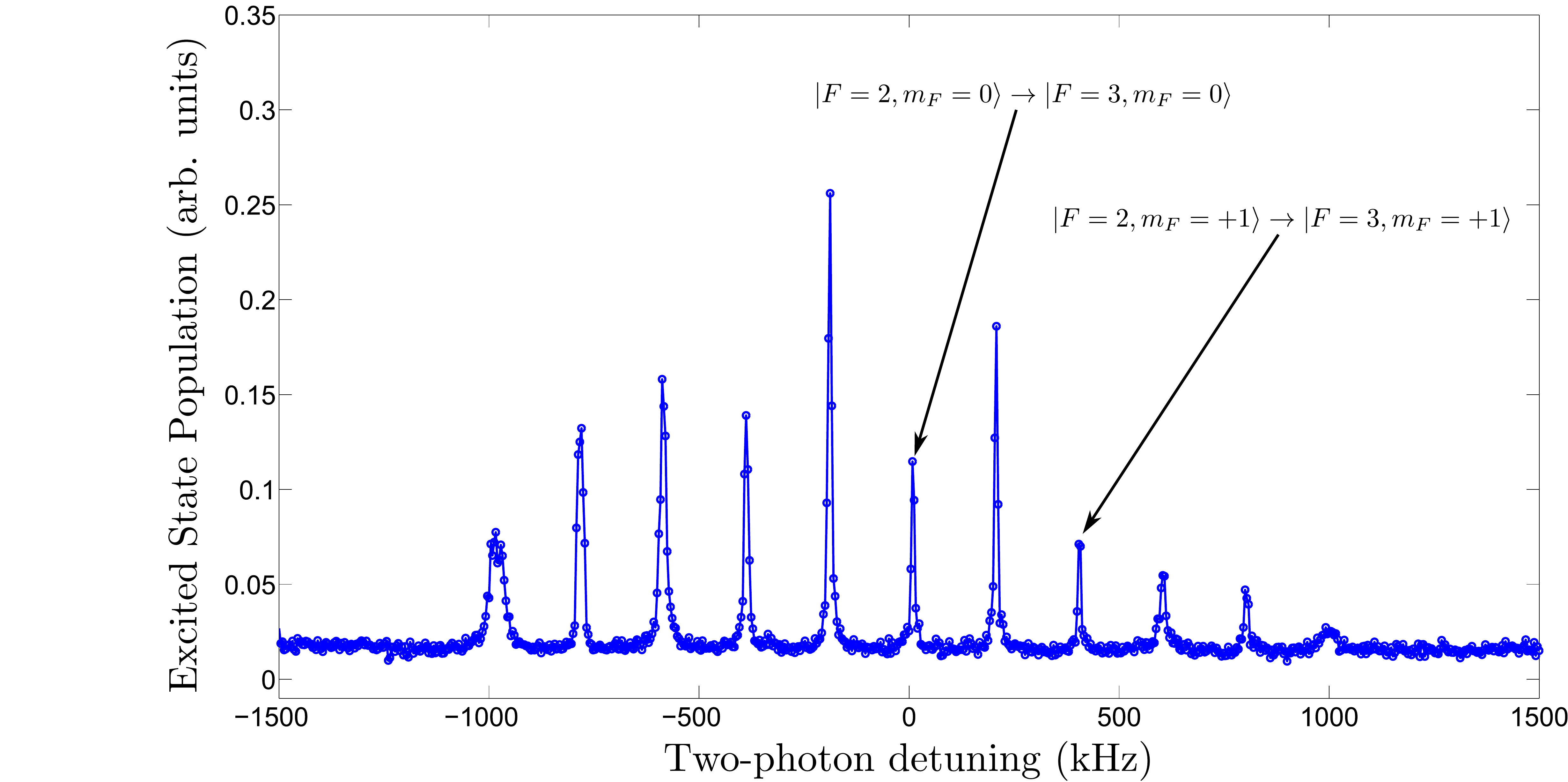}
}
\caption{Typical Raman spectrum containing the transitions between the states of the $F=2$ and $F=3$ manifolds for an arbitrarily aligned magnetic field
in their dependence on the two-photon detuning $\tilde{\Delta}$ defined by Eq. (\ref{Two-photon-detuning})
with respect to the clock transition, for which resonance occurs at $\tilde{\Delta}=0$.
We observe 11 resonances corresponding to 5 or 6 transitions with $\Delta m_F=0$ or $\Delta m_F=\pm 1$, respectively.
Transitions with $\Delta m_F=\pm 2$ are heavily suppressed \cite{Frank-JMO2013}.}
\label{fig: Raman spectrum}       
\end{figure*}

Using the $+2$ transition, that is $|F_1=2,m_{F_1}=1\rangle\to|F_2=3,m_{F_2}=1\rangle$, as indicated in Fig. \ref{fig: Raman spectrum},
we determine from the Zeeman shift, Eq. (\ref{detuning shift}), the magnetic field
\begin{equation}
 \label{magnetic field experiment}
    B(z)=0.11\tilde{\Delta}_{2,1}^{3,1}\,\frac{\mu {\rm T}}{{\rm kHz}},
\end{equation}
where we have used the fact \cite{Steck} that $m_{F_2}=m_{F_1}=1$ and $g_{F_2}=-g_{F_1}=1/3$.
It is the opposite signs of $g_{F_2}$ and $g_{F_1}$ that result in the non-degenerate spectrum.

By measuring the resonance frequencies corresponding to the clock and first (or second) peak in the Raman spectrum,
we automatically correct for any possible drift in the AC Stark shift caused by a drift in the intensity or frequency of the Raman fields.

We map out the magnetic field experienced by the atoms as a function of their location within the tower,
by repeated launching them and applying a Raman pulse at different times after their launch.
The corresponding location $z$ of the atom cloud is determined beforehand by time-of-flight photography.

\begin{figure*}
\centering
\resizebox{0.9\textwidth}{!}{%
  \includegraphics{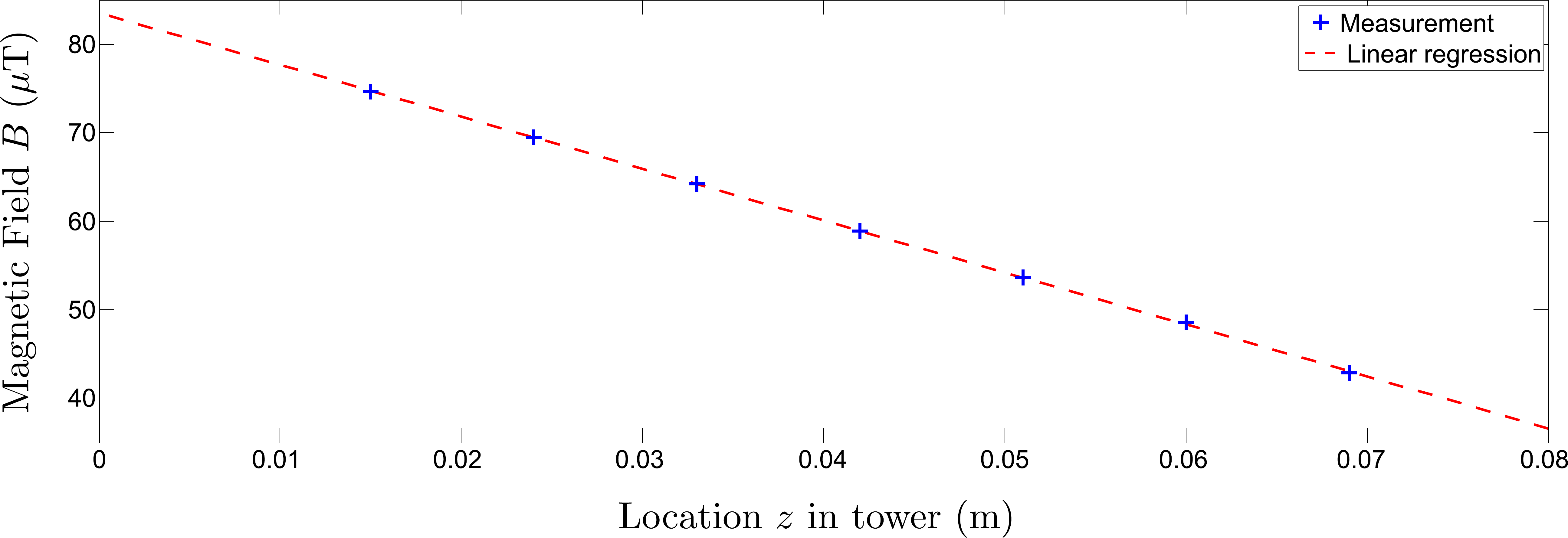}
}
\caption{Measurement of the magnetic field gradient along the vertical direction in the glass tower.
A linear regression of our data points deduced from the Raman spectra with the help of Eq. (\ref{magnetic field experiment})
yields the magnetic field $B(z)=(83.5-587{\rm m}^{-1}\cdot z)\,\mu \rm{T}$
along the tower and corresponds to a magnetic field gradient of approximately $600\,\mu{\rm T/m}$.
For this measurement we have used the frequency difference between the clock and the +2 transition induced by the Zeeman effect.
The magnetic field is generated by currents of $90\, {\rm mA}$ for the gradient coils and $40\, {\rm mA}$ for the solenoid.}
\label{fig: Magnetic field}       
\end{figure*}

Figure \ref{fig: Magnetic field} presents the measured magnetic field with a gradient of approximately $600\, \mu{\rm T/m}$
for a current of $90\, {\rm mA}$ in the gradient coils and $40\, {\rm mA}$ in the solenoid.


%


\subsection{Imperfect pulses and decoherence}
\label{subsec:4-4}

The theoretical proposal for a $T^3$-interferometer presented in the preceding sections assumes that the Raman pulses applied to the atoms
are perfect $\frac{\pi}{2}$- and $\pi$-pulses.
By definition, such pulses can occur only when the Raman fields are in a two-photon resonance.
Since this resonance shifts as the atoms travel up the tower, a rapid detuning of the relative frequency of the Raman fields is necessary.
We have been able to achieve this task using a combination of a high-frequency acousto-optic modulator and digital frequency synthesizer,
thereby keeping the atoms in resonance during their flight \cite{Frank-JMO2016}.

However, a more severe restriction is decoherence.
In order to observe the $T^3$-phase for a magnetic field gradient of $600\, \mu{\rm T/m}$, a time $T$ between pulses of about $1.5\, {\rm ms}$ is required.
Moreover, the decoherence time should be larger than $4T$, that is $6\, {\rm ms}$.
If we use the clock transition there is little to no decoherence on this time scale.
However, if we use a transition involving magnetically sensitive states, our signals decohere on the order of $200\, \mu{\rm s}$.
We are presently investigating the source of this decoherence.


%


\section{Summary and outlook}
\label{sec:6}
In the present article we have proposed an atom interferometer that is sensitive to the quantum mechanical $T^3$-phase emerging in the propagator
of a particle in a linear potential. For this purpose we have considered an atom with a magnetic sensitive, and an insensitive internal state being exposed
to a constant gravitational field and a magnetic field of constant gradient. By applying a sequence of four co-propagating Raman pulses,
the atom interferometer can be closed in position {\it and} velocity.
The resulting interferometer phase $\varphi_\mathrm{i}$ displays the cubic scaling in $T$ but also depends on
the gravitational acceleration and the magnetic field gradient.

We have compared and contrasted this cubic term to the one appearing in the phase of the Kasevich-Chu interferometer
in the presence of a gravity gradient, and to the one obtained by using the CAB technique.
Furthermore, we have outlined a possible experimental realization of our interferometer, which can be used as a gravimeter and magnetic gradiometer.

Cubic phases appear frequently in quantum physics and give rise to mind-boggling effects.
For example, the energy wave function of a linear potential is given by the Airy function \cite{Breit} whose standard integral representation
involves a cubic phase. This term emerges from the eigenvalue equation in momentum space due to the integration of the kinetic energy
which is quadratic in momentum.

When we suddenly turn-off the potential the so-created Airy wave packet accelerates and its probability density
keeps its shape \cite{Berry-Balazs} during the free propagation. Deeper insight \cite{Gravity-inertial mass}
into this surprising phenomenon springs from Wigner phase space \cite{Schleich-book} and the fact that
the Wigner function of the Airy wave packet is again an Airy function.

Closely related to the cubic phase in the Airy integral and the dispersionless free propagation of an Airy wave packet is the oscillatory probability density
created by a point source \cite{Berry} located in a linear potential and continuously emitting particles into all three space directions with an identical speed.
These oscillations appearing in the plane orthogonal to the gravitational force are a consequence of the interference between
two classical trajectories of different inclinations. The knowledge of the two distinct paths encoded in the different arrival times is erased by
the continuous stream of particles. Again the origin of this particular interference pattern can be traced back to the cubic phase in the Green's function.

Due to the analogy between the constant gravitational field and the constant electric field between two plates of a capacitor
discussed in the beginning of this article one might wonder if it is possible to construct a similar charged particle fountain.
Indeed, in the case of electrons in a uniform electric field such type of fountain has already been realized
in photoionization and photodetachment microscopes \cite{Yu,Kleber,Blonde}.

It would be fascinating to illuminate the similarities and differences between these three examples of cubic phases and our $T^3$-interferometer.
Unfortunately this task goes beyond the scope of the present article and has to be postponed to future publications.

\appendix

\section{Raman pulses: superpositions and exchanges}
\label{sec:A}

In this Appendix we describe the population dynamics \cite{Kasevich-Chu-1,Kasevich-Chu-2,Schleich-Greenberger-Rasel}
of the two resonant atomic states driven by the Raman laser pulses.
For this purpose we consider the interaction between a three-level atom and two laser pulses of the form
$$
{\bf E}_{1}(z,t)\equiv {\boldsymbol{\mathcal E}}_{1}(t)\cos\left(k_1 z-\omega_1t+\phi_1\right)
$$
and
\begin{equation}
 \label{Electric field}
  {\bf E}_2(z,t)\equiv{\boldsymbol{\mathcal E}}_{2}(t)\cos\left(k_2 z-\omega_2 t+\phi_2\right),
\end{equation}
where ${\boldsymbol{\mathcal E}}_j$, $k_j$, $\omega_j$, and $\phi_j$ with $j=1,2$ denote the time-dependent envelope, frequency, wave vector, and
phase of the $j$-th field, respectively.

The laser frequencies $\omega_1$ and $\omega_2$ are assumed to only drive the transitions
$|g_1\rangle \longleftrightarrow |e\rangle$ and $|g_2\rangle \longleftrightarrow |e\rangle$, respectively.
Moreover, we assume that the laser pulses are so short that the atom does not move significantly
during the interaction. Therefore, the position of the center-of-mass of the atom is considered to be fixed during the laser pulses.

Within the rotating-wave approximation \cite{Schleich-book} and in the limit of far-detuned laser pulses with identical detunings, that is
when the Rabi frequencies $\Omega_{j}(t)\equiv{\bf d}_{g_j e}\cdot{\boldsymbol{\mathcal E}}_j(t)/\hbar$ of the transitions
$|g_j\rangle \longleftrightarrow |e\rangle$ are much smaller than the detuning $\Delta_j\equiv \omega_j-\omega_{eg_j}$ of the two laser pulses,
$|\Delta_j|\gg |\Omega_j|$, and $\Delta\equiv\Delta_1=\Delta_2$, we can eliminate the excited state $|e\rangle$ and neglect
the Stark shifts $|\Omega_{j}(t)|^2/(4\Delta)$. The resulting effective Hamiltonian \cite{Schleich-Greenberger-Rasel}
\begin{equation}
 \label{Atom-light-Hamiltonian}
\begin{split}& \hat{H}_{\mathrm{p}}=\hbar\frac{\Omega_1(t)\Omega_2(t)}{4\Delta}\left(e^{i[\Delta k z+\phi_\mathrm{L}]}|g_1\rangle\langle g_2|\right. \\
& \left.+ e^{-i[\Delta k z+\phi_\mathrm{L}]}|g_2\rangle\langle g_1|\right)\end{split}
\end{equation}
describes the transitions between the states $|g_1\rangle$ and $|g_2\rangle$ due to the Raman pulses.
Here ${\bf d}_{g_j e}\equiv \langle g_j|{\bf d}|e\rangle$ and $\omega_{eg_j}\equiv (E_e-E_{g_j})/\hbar$
are the dipole-moment matrix element and the frequency of the transition $|g_j\rangle \longleftrightarrow |e\rangle$, respectively, with
$\Delta k\equiv k_2-k_1$ and the slowly varying laser phase $\phi_\mathrm{L}(t)\equiv \phi_2-\phi_1$.

To avoid a momentum transfer during the Raman transitions, we assume that the laser pulses propagate in the same directions
along the $z$-axis, Eq. (\ref{Electric field}), and that
the difference $\Delta k$ of the two wave vectors is small compared to the size $\delta z$ of the atomic wave packet, that is
$|\Delta k| \delta z\ll 1$. In this case the dependence on $z$ in Eq. (\ref{Atom-light-Hamiltonian}) can be neglected and we arrive at
\begin{equation}
 \label{Atom-light-Hamiltonian-effective}
  \hat{H}_{\mathrm{p}}\cong \hbar\frac{\Omega_1(t)\Omega_2(t)}{4\Delta}\left(e^{i\phi_\mathrm{L}}|g_1\rangle\langle g_2|+
    e^{-i\phi_\mathrm{L}}|g_2\rangle\langle g_1|\right).
\end{equation}

The interaction of the atom with the two far-detuned Raman pulses, corresponding to Eq. (\ref{Electric field}),
during the time interval $t_\mathrm{i}<t<t_\mathrm{f}$, and with $\Omega_j(t_{\mathrm{i}})=\Omega_j(t_{\mathrm{f}})=0$,
is given by the evolution operator \cite{Schleich-Greenberger-Rasel}
\begin{equation}
 \label{Atom-light-evaluation-definition}
 \begin{split}   &\hat{U}_\mathrm{p}\equiv \mathds{1}+\left(-\frac{i}{\hbar}\right)\int\limits_{t_\mathrm{i}}^{t_\mathrm{f}}dt \hat{H}_{\mathrm{p}}(t)\\
 &+ \left(-\frac{i}{\hbar}\right)^2\int\limits_{t_\mathrm{i}}^{t_\mathrm{f}}dt\int\limits_{t_\mathrm{i}}^{t}dt'\hat{H}_{\mathrm{p}}(t)\hat{H}_{\mathrm{p}}(t')+...\;, \end{split}
\end{equation}
which can be expressed as
\begin{equation}
 \label{Atom-light-evaluation-2}
\begin{split}&  \hat{U}_\mathrm{p}(\theta)=\left(\,|g_1\rangle\langle g_1|+|g_2\rangle\langle g_2|\,\right)\cos\left(\frac{\theta}{2}\right)\\\ &-
  i\left(e^{i\phi_\mathrm{L}}|g_1\rangle\langle g_2|+e^{-i\phi_\mathrm{L}}|g_2\rangle\langle g_1|\right)\sin\left(\frac{\theta}{2}\right),\end{split}
\end{equation}
where
\begin{equation}
 \label{pulse-area}
  \theta\equiv \frac{1}{2\Delta}\int\limits_{t_\mathrm{i}}^{t_\mathrm{f}}dt\,\Omega_1(t)\Omega_2(t)
\end{equation}
denotes the total pulse area.

The case $\theta=\frac{\pi}{2}$, which is a $\frac{\pi}{2}$-pulse, gives rise \cite{Giese} to the coherent superpositions
$$
\hat{U}_\mathrm{p}\left(\frac{\pi}{4}\right)|g_1\rangle=\frac{1}{\sqrt{2}}\left(|g_1\rangle-
  ie^{-i\phi_\mathrm{L}}|g_2\rangle \right)
$$
and
\begin{equation}
 \label{beam-splitter}
   \hat{U}_\mathrm{p}\left(\frac{\pi}{4}\right)|g_2\rangle=
  \frac{1}{\sqrt{2}}\left(|g_2\rangle-i e^{i\phi_\mathrm{L}}|g_1\rangle\right).
\end{equation}

In contrast, the case $\theta=\pi$, known as a $\pi$-pulse describes an exchange
$$
\hat{U}_\mathrm{p}\left(\frac{\pi}{2}\right)|g_1\rangle=-ie^{-i\phi_\mathrm{L}}|g_2\rangle
$$
and
\begin{equation}
 \label{pi-pulse}
  \hat{U}_\mathrm{p}\left(\frac{\pi}{2}\right)|g_2\rangle=-ie^{i\phi_\mathrm{L}}|g_1\rangle
\end{equation}
of the level populations.

\section{Interferometer: Sequence of unitary operators}
\label{sec:B}

Unitary operators describe both the interaction of the atom with the four Raman pulses and
the time evolution associated with the center-of-mass motion. In the present
Appendix we derive the complete quantum state of the atom
consisting of the internal states as well as the center-of-mass in the two exit ports of our interferometer
following the procedure outlined in Refs. \cite{Giese,Schleich-Greenberger-Rasel,Kleinert}.

The dynamics in our interferometer consists of the following steps:

1. Before the first $\frac{\pi}{2}$-pulse, at $t=t_0-\varepsilon$, the initial state
\begin{equation}
 \label{initial state}
  |\Psi(t_0-\varepsilon)\rangle\equiv |g_1\rangle|\psi_0\rangle
\end{equation}
consists of the center-of-mass motion $|\psi_0\rangle$ and the internal state $|g_1\rangle$.
Here and throughout this Appendix $\varepsilon$ is an infinitesimally small and positive number.

2. After the first $\frac{\pi}{2}$-pulse at $t=t_0+\varepsilon$, the state reads
\begin{equation}
 \label{first pulse}
 \begin{split}
  &|\Psi(t_0+\varepsilon)\rangle=\hat{U}_\mathrm{p}\left(\frac{\pi}{4}\right)|\Psi(t_0-\varepsilon)\rangle\\
  &=\left(\frac{1}{\sqrt{2}}|g_1\rangle-\frac{i}{\sqrt{2}}e^{-i\phi_\mathrm{L}(t_0)}|g_2\rangle\right)|\psi_0\rangle,\end{split}
\end{equation}
where we have used Eq. (\ref{beam-splitter}).

3. Before the first $\pi$-pulse at $t=t_1-\varepsilon$, we find
\begin{equation}
 \label{first gravity}
 \begin{split}& |\Psi(t_1-\varepsilon)\rangle=\hat{U}_a\left(t_1,t_0\right)|\Psi(t_0+\varepsilon)\rangle\\
 &= \frac{1}{\sqrt{2}}|g_1\rangle\hat{U}_{a_1}(t_1,t_0)|\psi_0\rangle
 \\ &- \frac{i}{\sqrt{2}}e^{-i\phi_\mathrm{L}(t_0)}|g_2\rangle\hat{U}_{a_2}(t_1,t_0)|\psi_0\rangle. \end{split}
\end{equation}

4. After the first $\pi$-pulse at $t=t_1+\varepsilon$, we obtain
\begin{equation}
 \label{second pulse}
 \begin{split}
 &|\Psi(t_1+\varepsilon)\rangle=\hat{U}_\mathrm{p}\left(\frac{\pi}{2}\right)|\Psi(t_1-\varepsilon)\rangle\\
  &=-\frac{i}{\sqrt{2}}e^{-i\phi_\mathrm{L}(t_1)}|g_2\rangle\hat{U}_{a_1}(t_1,t_0)|\psi_0\rangle
  \\& -\frac{1}{\sqrt{2}}e^{-i[\phi_\mathrm{L}(t_0)-\phi_\mathrm{L}(t_1)]}|g_1\rangle\hat{U}_{a_2}(t_1,t_0)|\psi_0\rangle.
  \end{split}
\end{equation}

5. Before the second $\pi$-pulse at $t=t_2-\varepsilon$, the state takes the form
\begin{equation}
 \label{second gravity}
\begin{split}
&|\Psi(t_2-\varepsilon)\rangle=\hat{U}_a(t_2,t_1)|\Psi(t_1+\varepsilon)\rangle\\
&=-\frac{i}{\sqrt{2}}e^{-i\phi_\mathrm{L}(t_1)}|g_2\rangle\hat{U}_{a_2}(t_2,t_1)\hat{U}_{a_1}(t_1,t_0)|\psi_0\rangle\\
&- \frac{1}{\sqrt{2}}e^{-i[\phi_\mathrm{L}(t_0)-\phi_\mathrm{L}(t_1)]}|g_1\rangle\hat{U}_{a_1}(t_2,t_1)\hat{U}_{a_2}(t_1,t_0)|\psi_0\rangle.\end{split}
\end{equation}

6. After the second $\pi$-pulse at $t=t_2+\varepsilon$, we arrive at the state
\begin{equation}
 \label{third pulse}
 \begin{split}
& |\Psi(t_2+\varepsilon)\rangle=\hat{U}_\mathrm{p}\left(\frac{\pi}{2}\right)|\Psi(t_2-\varepsilon)\rangle\\
&=-\frac{1}{\sqrt{2}}e^{-i[\phi_\mathrm{L}(t_1)-\phi_\mathrm{L}(t_2)]}|g_1\rangle\hat{U}_{a_2}(t_2,t_1)\hat{U}_{a_1}(t_1,0)|\psi_0\rangle
\\&+\frac{i}{\sqrt{2}}e^{-i[\phi_\mathrm{L}(t_0)-\phi_\mathrm{L}(t_1)+\phi_\mathrm{L}(t_2)]}\\
&\times|g_2\rangle\hat{U}_{a_1}(t_2,t_1)\hat{U}_{a_2}(t_1,t_0)|\psi_0\rangle.
  \end{split}
\end{equation}

7. Before the second $\frac{\pi}{2}$-pulse at $t=t_3-\varepsilon$, the state reads
\begin{equation}
 \label{third gravity}
\begin{split}
&|\Psi(t_3-\varepsilon)\rangle=\hat{U}_a(t_3,t_2)|\Psi(t_2+\varepsilon)\rangle\\
&=-\frac{1}{\sqrt{2}}e^{-i[\phi_\mathrm{L}(t_1)-\phi_\mathrm{L}(t_2)]}\\\
&\times|g_1\rangle\hat{U}_{a_1}(t_3,t_2)\hat{U}_{a_2}(t_2,t_1)\hat{U}_{a_1}(t_1,t_0)|\psi_0\rangle\\
&+ \frac{i}{\sqrt{2}}e^{-i[\phi_\mathrm{L}(t_0)-\phi_\mathrm{L}(t_1)+
  \phi_\mathrm{L}(t_2)]}\\
  &\times|g_2\rangle\hat{U}_{a_2}(t_3,t_2)\hat{U}_{a_1}(t_2,t_1)\hat{U}_{a_2}(t_1,t_0)|\psi_0\rangle.\end{split}
\end{equation}

8. Finally, after the second $\frac{\pi}{2}$-pulse at $t=t_3+\varepsilon$, we conclude with the state
\begin{equation}
 \label{fourth pulse}
 \begin{split}
& |\Psi(t_3+\varepsilon)\rangle=\hat{U}_\mathrm{p}\left(\frac{\pi}{4}\right)|\Psi(t_3-\varepsilon)\rangle\\
& =  \frac{1}{2}\,e^{-i[\phi_\mathrm{L}(t_1)-\phi_\mathrm{L}(t_2)]}|g_1\rangle\left(e^{-i\varphi_\mathrm{L}}\hat{U}_{\mathrm{u}}-\hat{U}_{\mathrm{l}}\right)|\psi_0\rangle\\
&+\frac{i}{2}e^{-i[\phi_\mathrm{L}(t_1)-\phi_\mathrm{L}(t_2)+\phi_\mathrm{L}(t_3)]}|g_2\rangle
  \left(e^{-i\varphi_\mathrm{L}}\hat{U}_{\mathrm{u}}+\hat{U}_{\mathrm{l}}\right)|\psi_0\rangle,\end{split}
\end{equation}
where
$$
\hat{U}_{\mathrm{l}}\equiv\hat{U}_{a_1}(t_3,t_2)\hat{U}_{a_2}(t_2,t_1)\hat{U}_{a_1}(t_1,t_0)
$$
and
\begin{equation}
 \label{U_l-U_u}
 \hat{U}_{\mathrm{u}}\equiv\hat{U}_{a_2}(t_3,t_2)\hat{U}_{a_1}(t_2,t_1)\hat{U}_{a_2}(t_1,t_0)
\end{equation}
are the unitary evolution operators associated with the center-of-mass motion for the lower and the upper paths
of the interferometer shown in Fig. \ref{fig: scheme}, and
\begin{equation}
 \label{laser phase}
  \varphi_\mathrm{L}\equiv \phi_\mathrm{L}(t_0)-2\phi_\mathrm{L}(t_1)+2\phi_\mathrm{L}(t_2)-\phi_\mathrm{L}(t_3)
\end{equation}
is the total phase resulting from the action of the four laser pulses.

\section{Conditions for a closed $T^3$-interferometer}
\label{sec:C}

In the preceding Appendix we have derived an expression for the complete quantum state of the atom in the exit ports of the interferometer.
Here we have allowed arbitrary times for the interactions with the laser pulses. In the present Appendix we choose these times in such a way
as to maximize the contrast.

The probability $P_{g_1}$ to observe atoms in the ground state $|g_1\rangle$ after the action of the four Raman pulses at $t=t_3+\varepsilon$, follows from the quantum state $|\Psi(t_3+\varepsilon)\rangle$ given by Eq. (\ref{fourth pulse})
and contains the state
\begin{equation}
 \label{interferometer-psig1}
  |\psi_{g_1}\rangle\equiv\langle g_1|\Psi(t_3+\varepsilon)\rangle
\end{equation}
of the center-of-mass motion of atom in $|g_1\rangle$. It takes the form
\begin{equation}
 \label{interferometer-probability}
  P_{g_1}\equiv\langle\psi_{g_1}|\psi_{g_1}\rangle=\frac{1}{2}\left[1-C\cos\left(\varphi_\mathrm{i}+\varphi_\mathrm{L}\right)\right],
\end{equation}
where the contrast $C$ and the phase $\varphi_\mathrm{i}$ of the interferometer are the modulus and the argument of the matrix element
\begin{equation}
 \label{C-phi_i}
  \langle \psi_0|\hat{U}^{\dagger}_\mathrm{u} \hat{U}^{}_\mathrm{l}|\psi_0\rangle\equiv Ce^{i\varphi_\mathrm{i}}.
\end{equation}
We maximize $C$, that is we have $C=1$, when we close our interferometer.
In this case $P_{g_1}$ given by Eq. (\ref{interferometer-probability}) is independent of
the initial velocity and position of the atom.

In order to close the interferometer we have to find the time intervals $t_{j+1,j}\equiv t_{j+1}-t_{j}$ with $j=0,1,2$
between the Raman pulses shown in Fig. \ref{fig: scheme},
such that the final velocities $v_\mathrm{u}(t_3)$ and $v_\mathrm{l}(t_3)$, as well as the final positions $z_\mathrm{u}(t_3)$ and $z_\mathrm{l}(t_3)$
on the upper and lower paths of the interferometer are identical.

Indeed, for the velocity we derive the following formulae:

i) for the upper path
\begin{equation*}
\begin{split}v_0&\rightarrow v_\mathrm{u}(t_1)=v_0+a_2t_{10}\\
&\rightarrow v_\mathrm{u}(t_2)=v_\mathrm{u}(t_1)+a_1t_{21}\\
&\rightarrow v_\mathrm{u}(t_3)=v_\mathrm{u}(t_2)+a_2t_{32}\end{split}
\end{equation*}

ii) for the lower path
\begin{equation*}
\begin{split}v_0&\rightarrow v_\mathrm{l}(t_1)=v_0+a_1t_{10}\\
&\rightarrow v_\mathrm{l}(t_2)=v_\mathrm{l}(t_1)+a_2t_{21}\\
&\rightarrow v_\mathrm{l}(t_3)=v_\mathrm{l}(t_2)+a_1t_{32}.\end{split}
\end{equation*}

\noindent As a result, the interferometer is closed in {\it velocity} space, if $v_ \mathrm{u}(t_3)=v_\mathrm{l}(t_3)$, that is,
\begin{equation*}
\begin{split}&v_0+a_2t_{10}+a_1t_{21}+a_2t_{32}\\
=&v_0+a_1t_{10}+a_2t_{21}+a_1t_{32},\end{split}
\end{equation*}
or, equivalently,
\begin{equation}
\label{v-condition}
    t_{10}-t_{21}+t_{32}=0.
\end{equation}

As for the position, we obtain the following rather lengthy expressions:

i) for the upper path
\begin{equation*}
\begin{split}z_0&\rightarrow z_\mathrm{u}(t_1)=z_0+v_0t_{10}+\frac{1}{2}a_2 t_{10}^2\\
&\rightarrow z_\mathrm{u}(t_2)=z_\mathrm{u}(t_1)+v_\mathrm{u}(t_1)t_{21}+\frac{1}{2}a_1 t_{21}^2\\
&\rightarrow z_\mathrm{u}(t_3)=z_\mathrm{u}(t_2)+v_\mathrm{u}(t_2)t_{32}+\frac{1}{2}a_2 t_{32}^2\\
&=z_0+v_0(t_{10}+t_{21}+t_{32})+\frac{1}{2}(a_2t_{10}^2+a_1t_{21}^2+a_2t_{32}^2)\\&+a_2t_{10}(t_{21}+t_{32})+a_1t_{21}t_{32},
\end{split}
\end{equation*}

ii) for the lower path
\begin{equation*}
\begin{split}
z_0&\rightarrow z_\mathrm{l}(t_1)=z_0+v_0t_{10}+\frac{1}{2}a_1 t_{10}^2\\
&\rightarrow z_\mathrm{l}(t_2)=z_\mathrm{l}(t_1)+v_\mathrm{l}(t_1)t_{21}+\frac{1}{2}a_2 t_{21}^2\\
&\rightarrow z_\mathrm{l}(t_3)=z_\mathrm{l}(t_2)+v_\mathrm{l}(t_2)t_{32}+\frac{1}{2}a_1 t_{32}^2\\
&=z_0+v_0(t_{10}+t_{21}+t_{32})+\frac{1}{2}(a_1t_{10}^2+a_2t_{21}^2+a_1t_{32}^2)\\
&+a_1t_{10}(t_{21}+t_{32})+a_2t_{21}t_{32}.
\end{split}
\end{equation*}

\noindent As a result, the interferometer is closed in {\it position} space if $z_\mathrm{u}(t_3)=z_\mathrm{l}(t_3)$, that is,
\begin{equation}
 \label{z-condition}
  t_{10}^2-t_{21}^2+t_{32}^2+2t_{10}(t_{21}+t_{32})-2t_{21}t_{32}=0.
\end{equation}

When we solve the system of the two algebraic equations, (\ref{v-condition}) and (\ref{z-condition}), for $t_{21}$ and $t_{32}$
in terms of $t_{10}$, we obtain
\begin{equation}
 \label{t-condition}
  t_3-t_2=t_1-t_0=T\;\;\;{\rm and}\;\;\;t_2-t_1=2T.
\end{equation}

Hence, in order to close the interferometer, the four Raman pulses must be separated in time by $T$, $2T$, and $T$ as indicated in Fig. \ref{fig: scheme}.

\section{Interferometer phase}
\label{sec:D}

In the preceding Appendix we have used classical trajectories to find the separation $T-2T-T$ between the four Raman pulses
leading to a closed interferometer. We now show that in this case 
the product $\hat{U}_\mathrm{u}^{\dag}\hat{U}_\mathrm{l}$ of the evolution operators
$\hat{U}_\mathrm{l}$ and $\hat{U}_\mathrm{u}$ defined by Eq. (\ref{U_l-U_u})
is proportional \cite{Giese,Roura2014,Kleinert} to the identity operator, that is
\begin{equation}
 \label{interferometer phase-operator product}
  \hat{U}_\mathrm{u}^{\dag}\hat{U}^{}_\mathrm{l}=e^{i\varphi_\mathrm{i}}\,\mathds{1},
\end{equation}
where $\varphi_{\rm i}$ is the interferometer phase.

Therefore, a normalized state $|\psi_0\rangle$ leads by virtue of  Eq. (\ref{C-phi_i}) to a perfect contrast, that is $C=1$, 
indicating that the interferometer is independent of $|\psi_0\rangle$. 
Moreover, this calculation provides us with an explicit expression for $\varphi_{\rm i}$.

In order to evaluate the evolution operator
\begin{equation}
 \label{U_l product}
  \hat{U}^{}_{\rm l}\equiv\hat{U}_{a_1}(T)\hat{U}_{a_2}(2T)\hat{U}_{a_1}(T)
\end{equation}
for the lower path of our interferometer, shown in Fig. \ref{fig: scheme},
we use the Baker-Campbell-Hausdorff and Zassenhaus formulas \cite{Wilcox} to represent the operator $\hat{U}_{a}(T)$ given by Eq. (\ref{operator-U-gravity})
in the form of a product
\begin{equation}
 \label{BCH}
  \hat{U}_{a}(T)=\exp\left(i\frac{ma^2T^3}{12\hbar}\right)\hat{\mathcal{D}}\left(\frac{1}{2}aT^2,maT\right)\hat{U}_{0}(T)
\end{equation}
consisting of a phase factor, the displacement operator
\begin{equation}
 \label{BCH-displacement operator}
  \hat{\mathcal{D}}\left(Z,P\right)\equiv\exp\left[-\frac{i}{\hbar}\left(Z\hat{p}_z-P\hat{z}\right)\right],
\end{equation}
and the unitary operator
\begin{equation}
 \label{BCH-free evolution}
  \hat{U}_{0}(T)\equiv\exp\left(-\frac{i}{\hbar}\frac{\hat{p}_z^2}{2m}T\right)
\end{equation}
of a free particle.

The decomposition, Eq. (\ref{BCH}), allows us to rewrite Eq. (\ref{U_l product}) as
$$
\hat{U}^{}_\mathrm{l}=\exp\left[i\frac{m(a_1^2+4a_2^2)}{6\hbar}T^3\right]
\hat{\mathcal{D}}\left(\frac{1}{2}a_1 T^2,ma_1T\right)\hat{U}_{0}(T)
$$
\begin{equation}
 \label{U_l product total}
  \times \hat{\mathcal{D}}\left(2a_2 T^2,2ma_2T\right)\hat{U}_{0}(2T)
\hat{\mathcal{D}}\left(\frac{1}{2}a_1 T^2,ma_1T\right)\hat{U}_{0}(T).
\end{equation}

With the help of the commutation relation
$$
\hat{U}_{0}(T)\hat{\mathcal{D}}\left(Z,P\right)=\hat{\mathcal{D}}\left(Z+\frac{P}{m}T,P\right)\hat{U}_{0}(T)
$$
and the addition identity
$$
\hat{U}_{0}(T_1)\hat{U}_{0}(T_2)=\hat{U}_{0}(T_1+T_2)
$$
for the operators $\hat{\mathcal{D}}$ and $\hat{U}$ given by Eqs. (\ref{BCH-displacement operator}) and (\ref{BCH-free evolution}),
we can shift all free-evolution operators $\hat{U}_0$ in Eq. (\ref{U_l product total}) to the right and we arrive at
$$
\hat{U}^{}_\mathrm{l}=\exp\left[i\frac{m(a_1^2+4a_2^2)}{6\hbar}T^3\right]
\hat{\mathcal{D}}\left(\frac{1}{2}a_1 T^2,ma_1T\right)
$$
$$
\times \hat{\mathcal{D}}\left(4a_2 T^2,2ma_2T\right)
\hat{\mathcal{D}}\left(\frac{7}{2}a_1 T^2,ma_1T\right)\hat{U}_{0}(4T),
$$
or
$$
\hat{U}^{}_\mathrm{l}=\exp\left[i\frac{m T^3}{3\hbar}\left(5a_1^2+9a_1a_2+2a_2^2\right)\right]
$$
\begin{equation}
 \label{U_l product final}
    \times\hat{\mathcal{D}}\left(4(a_1+a_2)T^2,2m(a_1+a_2)T\right)\hat{U}_{0}(4T).
\end{equation}

In the last step we have made use the addition identity 
$$
\hat{\mathcal{D}}\left(Z_1,P_1\right)\hat{\mathcal{D}}\left(Z_2,P_2\right)=e^{i\tilde{\varphi}}
\hat{\mathcal{D}}\left(Z_1+Z_2,P_1+P_2\right)
$$
with 
$$
\tilde{\varphi}\equiv\frac{1}{2\hbar}\left(P_1Z_2-P_2 Z_1\right),
$$
to combine all three displacement operators into a single one.

Since the evolution operator $\hat{U}^{}_\mathrm{u}$ defined by Eq. (\ref{U_l-U_u}) for the upper path of our interferometer
follows directly from the operator $\hat{U}^{}_\mathrm{l}$ given by Eq. (\ref{U_l product final}) for the lower path 
by an exchange of the accelerations $a_1$ and $a_2$, we arrive at
$$
\hat{U}^{}_\mathrm{u}=\exp\left[i\frac{m T^3}{3\hbar}\left(2a_1^2+9a_1a_2+5a_2^2\right)\right]
$$
\begin{equation}
 \label{U_u product final}
    \times\hat{\mathcal{D}}\left(4(a_1+a_2)T^2,2m(a_1+a_2)T\right)\hat{U}_{0}(4T).
\end{equation}

When we substitute Eqs. (\ref{U_l product final}) and (\ref{U_u product final}) into the left-hand side of Eq. (\ref{interferometer phase-operator product})
and use the property that the operators $\hat{\mathcal{D}}$ and $\hat{U}_{0}$ are unitary,
the interferometer phase reads
\begin{equation}
 \label{interferometer phase-result}
  \varphi_\mathrm{i}=\frac{m}{\hbar}\left(a_1^2-a_2^2\right)T^3.
\end{equation}

Hence, $\varphi_{\rm i}$ is independent of the initial position $z_0$ and velocity $v_0$ as well as of the initial state. 
Moreover, it scales with the third power of the time interval $T\equiv t_1-t_0$ between the first and the second Raman pulses.

\section{Alternative derivation of interferometer phase}
\label{sec:E}

In this appendix we provide an alternative derivation of the phase shift for our $T^3$-interferometer 
by making use of the general results obtained in Ref. \cite{Roura2014}. 
Here both open and closed atom interferometers with branch-dependent forces were investigated. 
We first summarize the phase accumulated along a single trajectory and then consider the interferometer, 
that is  we analyze the phase difference between two different paths.

\subsection{Phases accumulated during motion}

The framework outlined in Ref. \cite{Roura2014} describes light-pulse atom interferometers 
where the time evolution of the center-of-mass motion between laser pulses is governed by a general quadratic Hamiltonian and 
the time dependence of the corresponding quantum state along each interferometer branch is given by
\begin{equation}
|\psi (t)\rangle = e^{i \Phi(t)}\, \hat{\mathcal{D}} \big( \boldsymbol{\chi}(t)  \big) \, | \psi_\text{c} (t) \rangle
\label{eq:state_evol}.
\end{equation}
Here $|\psi_{\rm c}(t)\rangle$ denotes a {\it centered} state with vanishing expectation values for the position and momentum operators 
and evolves according to the purely quadratic part of the Hamiltonian.

It is convenient to employ a vector notation for phase-space quantities and the argument of the displacement operator 
$\hat{\mathcal{D}}$ is a displacement vector $\boldsymbol{\chi}(t)\equiv\big(\boldsymbol{\mathcal{R}}(t), \boldsymbol{\mathcal{P}}(t) \big)^\text{T}$ 
that corresponds to the classical phase-space trajectories 
$\boldsymbol{\mathcal{R}}=\boldsymbol{\mathcal{R}}(t)$ and $\boldsymbol{\mathcal{P}}=\boldsymbol{\mathcal{P}}(t)$
associated with the Hamiltonian including the momentum kicks from the laser pulses along that branch. 
Its value $\boldsymbol{\chi} (t_0)\equiv\boldsymbol{\chi}_0$ at the initial time $t_0$ coincides with 
the expectation values of the position and momentum operators for the initial state of the interferometer.

The time-dependent phase $\Phi$ reads
\begin{equation}
 \label{eq:total_phase}
  \Phi\equiv\varphi-\frac{1}{2\hbar}\int\limits_{t_0}^t dt' 
\left\{\left[\boldsymbol{\mathcal{F}}_\text{lp}^\text{T}(t') + \boldsymbol{\mathcal{G}}^\text{T}(t')\right] J \boldsymbol{\chi}(t')+2V_0 (t')\right\},
\end{equation}
where $\varphi$ is the sum of the various phases for the laser pulses relevant to this branch and $\boldsymbol{\mathcal{F}}_\text{lp}$ 
accounts for their momentum kicks.

The quantity $\boldsymbol{\mathcal{G}}$ is a consequence of the linear terms in the Hamiltonian and reduces 
for the case of a constant force considered in the present article to
\begin{equation}
\boldsymbol{\mathcal{G}}(t) \equiv
\left( \begin{array}{c} \mathbf{0} \\ m \mathbf{g}(t) \end{array} \right).
\label{eq:g_force}
\end{equation}

In addition, we have introduced the symplectic form
\begin{equation}
J \equiv \left( \begin{array}{cc} 0 & \mathds{1} \\ -\mathds{1} & 0 \end{array} \right)
\label{eq:symplectic_form}.
\end{equation}

Finally the term with $V_0$ contains the contributions from the internal state energies and can also include the effect of uniform magnetic fields 
which can even be time-dependent due to the Zeeman effect.

Since a solution $\boldsymbol{\chi}=\boldsymbol{\chi}(t)$ of the classical equations of motion in phase space 
consists of a homogeneous solution containing the information on the initial conditions plus an inhomogeneous solution 
that accounts for the linear terms and the kicks from the laser pulses,  $\boldsymbol{\mathcal{G}}$ and $\boldsymbol{\mathcal{F}}_\text{lp}$, we find
\begin{equation}
 \label{eq:full_solution}
\boldsymbol{\chi} (t) = \mathcal{T} (t,t_0)\, \boldsymbol{\chi}_0
+ (\mathcal{T}_\text{ret} \cdot \boldsymbol{\mathcal{G}}) (t)
+ (\mathcal{T}_\text{ret} \cdot \boldsymbol{\mathcal{F}}_\text{lp}) (t).
\end{equation}
Here the transition matrix $\mathcal{T}(t,t_0)$ satisfies the homogeneous part of the equations of motion 
with the initial condition $\mathcal{T} (t_0,t_0) = \mathds{1}$. 
Moreover, we have employed the retarded propagator $\mathcal{T}_\text{ret} (t,t') \equiv \mathcal{T} (t,t') \, \theta(t-t')$ and introduced the notation
\begin{equation}
 \label{eq:convolution}
    (\mathcal{T}_\text{ret} \cdot \boldsymbol{\mathcal{A}})(t)\equiv \int\limits^t_{t_0} dt' \mathcal{T}_\text{ret} (t,t')\boldsymbol{\mathcal{A}}(t').
\end{equation}

When we restrict ourselves to the familiar kinetic term associated with the free evolution due to the purely quadratic part of the Hamiltonian, 
the transition matrix simplifies to
\begin{equation}
\mathcal{T} (t,t') =
\left( \begin{array}{cc}
\mathds{1} & \frac{(t-t')}{m}\, \mathds{1} \\
\\
0 & \mathds{1}
\end{array} \right)
\label{eq:transition_free}.
\end{equation}

\subsection{Interferometer phase}

The oscillations in the number of atoms detected at each exit port as a result of the interference 
between the upper and lower branches of the interferometer is determined by the phase shift
\begin{equation}
 \label{eq:phase_shift1}
    \delta\Phi\equiv\Phi_{\rm l} - \Phi_{\rm u} + \frac{1}{2 \hbar}\boldsymbol{\chi}_{\rm u}^\text{T} J \boldsymbol{\chi}_{\rm l}.
\end{equation}
The last term arises only in {\it open} interferometers, where the central position and momentum of the two interfering wave packets 
do not coincide, that is $\boldsymbol{\chi}_{\rm l} \neq \boldsymbol{\chi}_{\rm u}$. 

For the case of no momentum transfer from the laser pulses, that is $\boldsymbol{\mathcal{F}}_\text{lp} = \mathbf{0}$, the phase difference 
$\delta \Phi$ reduces to
\begin{align}
 \label{eq:phase_shift3}
\delta\Phi(t) & = \delta\varphi - \frac{1}{\hbar} \int\limits_{t_0}^t dt'\, \delta V_0 (t')
- \frac{1}{\hbar} \delta \boldsymbol{\chi}^\text{T}(t) J \mathcal{T}(t,t_0)\boldsymbol{\chi}_0
\nonumber \\
&\quad - \frac{1}{\hbar} \int\limits_{t_0}^t dt' \int\limits_{t_0}^{t'} dt'' \,
\delta\boldsymbol{\mathcal{G}}^\text{T}(t') \, J\, \mathcal{T}(t',t'')
\boldsymbol{\bar{\mathcal{G}}}(t'').
\end{align}
Here a bar over a quantity $\mathcal{A}$ denotes the average of its values in the two branches, that is 
$\bar{\mathcal{A}}\equiv (\mathcal{A}_{\rm l} + \mathcal{A}_{\rm u})/2$, and 
similarly their difference is $\delta\mathcal{A}\equiv \mathcal{A}_{\rm l}-\mathcal{A}_{\rm u}$ 
in all cases except for $\delta\Phi$, which is defined otherwise above.

The quantity $\delta \boldsymbol{\chi}$ corresponds to the relative displacement between 
the two interfering wave packets at the exit port. Since our $T^3$-interferometer is closed, the third term on the right-hand side 
of Eq. \eqref{eq:phase_shift3} vanishes.

Moreover, the atoms spend the same times in each internal state on both branches. As a consequence 
the second term on the right-hand side of Eq. \eqref{eq:phase_shift3} vanishes as well.

Therefore,  when we substitute Eqs. \eqref{eq:g_force} and \eqref{eq:transition_free} into Eq. \eqref{eq:phase_shift3}, we arrive at
\begin{equation}
 \label{eq:phase_shift4}
\delta\Phi(t) = \varphi_{\rm L}+\frac{m}{\hbar} \int\limits_{t_0}^t dt' \int\limits_{t_0}^{t'} dt'' 
\delta\mathbf{g}^\text{T}(t')(t'-t'')\mathbf{\bar{g}}(t''),
\end{equation}
where we have taken into account that $\delta\varphi=\varphi_{\rm L}$ in our case.

Between the first and the forth pulse, the internal states and the accelerations experienced by the atoms 
on the two branches are exchanged every time a pulse is applied. As a result, we find
\begin{equation}
 \label{eq:g_evol1}
    \mathbf{\bar{g}}(t)\equiv \frac{1}{2}(\mathbf{a}_1 + \mathbf{a}_2) = \mathrm{const}
\end{equation}
and 
\begin{equation}
 \label{eq:g_evol2}
    \delta\mathbf{g}(t)\equiv f(t)(\mathbf{a}_1 - \mathbf{a}_2) 
\end{equation}
with
\begin{equation}
\label{eq:f_evol}
f(t)\equiv \left\{
\begin{array}{l}
+1,\quad t_0 < t < t_0 + T \\
-1,\quad t_0 + T < t < t_0 + 3T \\
+1,\quad t_0 + 3T < t < t_0 + 4T.
\end{array}
\right.
\end{equation}

Since $\mathbf{\bar{g}}$ is time independent we can immediately integrate over $t''$ and obtain
\begin{equation}
    \delta\Phi = \varphi_{\rm L}
    + \frac{m}{4 \hbar} (\mathbf{a}_1^2 - \mathbf{a}_2^2) \int\limits_{t_0}^{t_0 + 4T}dt' f(t')(t'-t_0)^2
\end{equation}
or 
\begin{equation}
  \delta\Phi = \varphi_{\rm L} + \frac{m}{\hbar} (\mathbf{a}_1^2 - \mathbf{a}_2^2)\, T^3,
\end{equation}
which agrees with the result of Appendix \ref{sec:D}.

\section*{Acknowledments}
We are grateful to E. Giese, M. A. Kasevich, S. Kleinert, H. M\"uller, G. Welch, and  W. Zeller for many fruitful discussions on this topic.
Moreover, we thank N. Ashby for pointing out Ref. \cite{Kennard} to us.

This work is supported by DIP, the German-Israeli Project Cooperation, as well as the German Space Agency (DLR) with funds provided by
the Federal Ministry for Economic Affairs and Energy (BMWi) due to an enactment of the German Bundestag
under Grants No. DLR 50WM1152-1157 (QUANTUS-IV) and the Centre for Quantum Engineering and Space-Time Research QUEST.

We appreciate the funding by the German Research Foundation (DFG) in the framework of the SFB/TRR-21.
W.P.S. is grateful to Texas A$\&$M University for a Texas A$\&$M University Institute for Advanced Study (TIAS) Faculty Fellowship. S.A.D., J.P.D., A.S., and F.A.N.
gratefully acknowledge funding from the Office of Naval Research
and a grant from the Naval Air Systems Command Chief Technology Office.

%
%

\end{document}